\RequirePackage{fix-cm}
\documentclass[smallextended]{svjour3}       
\smartqed  
\usepackage{graphicx}
\usepackage{siunitx}
\usepackage{url}
\usepackage{hyperref}
\usepackage{enumitem}
\usepackage{wrapfig}
\usepackage[table]{xcolor}
\usepackage{tabularx}
\usepackage{float}
\usepackage{placeins}
\usepackage{capt-of}
\usepackage{soul}
\usepackage{natbib}
\usepackage{amsmath}

\usepackage{booktabs}
\usepackage{graphicx}
\usepackage{xcolor}
\usepackage{pifont}
\usepackage{threeparttable}
\usepackage{array}
\newcommand{\cmark}{\textcolor{green!60!black}{\ding{51}}}
\newcommand{\xmark}{\textcolor{red!70!black}{\ding{55}}}
\usepackage{makecell}

\newcommand{\rqonea} {\textbf{Outcomes (case study)}: In this classroom deployment, how do contrasting dyads with similar normalized pre--post learning gains differ in their in-environment task mastery, frequency of interaction with CoSMoAgent, peer collaboration, and problem-solving strategies?}
\newcommand{\rqtwo} {\textbf{Learning Processes}: How do students’ verbalized understanding in response to CoSMoAgent probes and their reliance-oriented help-seeking behaviors change as their mastery of a modeling task increases?}
\newcommand{\rqthree}{\textbf{Adaptive Support}: How does CoSMoAgent’s dialogue-policy selection change as students’ task mastery evolves?}

\newcommand{\LowEngage}{Validation Seekers}
\newcommand{\HighEngage}{Guidance Seekers}

\newcommand{\Title}{A Theory-Guided LLM Pedagogical Agent Supporting Open-Ended STEM+C Learning}

\begin{document}

\let\WriteBookmarks\relax

\title{\Title}

\titlerunning{Theory-Guided LLM Agent for Open-Ended STEM+C Learning}

\author{Clayton Cohn \and
        Surya Rayala \and
        Siyuan Guo \and
        Hanchen David Wang \and
        Naveeduddin Mohammed \and
        Umesh Timalsina \and
        Shruti Jain \and
        Ryan Li \and
        Angela Eeds \and
        Menton Deweese \and
        Pamela J. Osborn Popp \and
        Rebekah Stanton \and
        Shakeera L. Walker \and
        Ashwin T S \and
        Meiyi Ma \and
        Gautam Biswas}

\authorrunning{Cohn et al.}

\institute{C. Cohn \at
              Cornell Ann S. Bowers College of Computing and Information Science, National Tutoring Observatory, Future of Learning Lab, Cornell University, Ithaca, NY 14850, USA \\
              \email{clayton.cohn@cornell.edu}
           \and
           S. Rayala \and S. Guo \and H. D. Wang \and N. Mohammed \and U. Timalsina \and S. Jain \and R. Li \and A. Eeds \and M. Deweese \and P. J. Osborn Popp \and R. Stanton \and S. L. Walker \and A. T S \and M. Ma \and G. Biswas \at
              Vanderbilt University, Nashville, TN 37235, USA
}

\date{}

\maketitle

\begin{abstract}
Open-ended learning environments (OELEs) can support STEM+C (science, technology, engineering, mathematics, and computing) inquiry through activities such as computational modeling, which requires students to translate disciplinary knowledge across domains and coordinate scientific concepts with computational representations. However, this is often challenging for students, many of whom require support during problem solving. In this exploratory study, we present CoSMoAgent (\textbf{Co}mputational \textbf{S}cience \textbf{Mo}deling \textbf{Agent}), a GPT-5-based multi-agent knowledgeable peer for high-school kinematics modeling grounded in the Evidence–Decision–Feedback (EDF) framework. CoSMoAgent goes beyond dialogue: it infers students’ strategies, learner states, and knowledge gaps from environment logs and evolving model states. These inferences guide dialogue policies that prioritize probing and hints over direct answers. We deployed CoSMoAgent with 11 dyads across three modeling tasks (33 dyad-sessions), using a mixed-methods design to examine learning outcomes, processes, and adaptive scaffolding. As task mastery increased, students verbalized their understanding more successfully and sought help less often, and the agent’s policies shifted from probing toward suggesting actions and extending reasoning. Two contrasting case-study dyads followed markedly different problem-solving paths, yet achieved similar learning gains despite substantial differences in task mastery and CoSMoAgent interaction. Students valued the agent’s questions but were often frustrated by its withholding of answers. The study contributes a design pattern and classroom evidence for LLM scaffolding grounded in environmental activity rather than chat alone within an open-ended modeling environment. 

\end{abstract}

\section{Introduction}


Open-ended learning environments (OELEs) are grounded in constructivist and inquiry-based learning: students build understanding by generating and testing ideas and making sense of results, rather than by receiving and practicing predefined procedures \citep{land2000cognitive}. This approach is especially valuable for K–12 STEM+C learning, where students integrate scientific concepts with computational representations to build, test, and explain models of phenomena \citep{weintrop2016defining}. Doing so requires practices central to both science and computing, including decomposition, testing, diagnosing mismatches, and debugging errors that may arise from scientific reasoning, code, or both \citep{sengupta2013integrating}. The learning goals in OELEs therefore extend beyond producing a correct computational artifact: students must also develop the cognitive skills to interpret evidence, coordinate disciplinary and computational representations, test and revise models, and regulate their progress through problems with multiple viable solution paths \citep{cohn2026edf}.

Developing and applying problem-solving strategies is an intended learning outcome in OELEs \citep{land2000cognitive,sengupta2013integrating}. \textit{Completing a task is not the same as learning from it}. Students can produce a working model through trial and error or answer-seeking without understanding why it works, and can also learn substantially without reaching full task completion. They often struggle to connect their actions to underlying disciplinary ideas \citep{basu2015scaffolding,snyder2024analyzing} and may prioritize task scores over understanding, ``gaming'' the system in ways that create the appearance of mastery \citep{baker2004off,cohn2026edf}.

As a result, evaluating support in OELEs requires distinguishing three often-conflated concepts: \textit{task mastery} within the environment, \textit{learning gains} measured independently of the task, and students' \textit{verbalized understanding}. It also requires attending to how problem solving unfolds as temporal patterns of action, not only to the solutions students produce \citep{snyder2024understanding,winne2017cognition}. These same demands also shape the design of effective support: pedagogical agents must respond to students' evolving understanding and activity without reducing open-ended problem solving to task completion.

Large language model (LLM)-based pedagogical agents offer a promising mechanism for providing such support in OELEs, including support from a ``knowledgeable'' peer within the Zone of Proximal Development (ZPD; \citealp{vygotsky1978mind}). Yet while some recent agents build on established AIED and learning sciences traditions \citep{cohn2026edf}, others foreground architectural novelty more than the learning mechanisms they are meant to support \citep{li_edumas_2024,wang_llm-powered_2025,zha_mentigo_2025}, prompting calls to re-anchor LLM agents in learning theory \citep{stamper2024enhancing}.

Weakly-grounded support can promote cognitive offloading and undermine critical thinking \citep{kosmyna2025your,stadler2024cognitive}. Theory-guided constraints can reduce this risk (e.g., by not providing students with direct answers), but may frustrate or disengage students who expect the agent to do their work for them. We term this tension the \textit{helpfulness paradox} \citep{cohn2026edf}: agents must provide enough support to sustain engagement while preserving the reasoning students are meant to perform. How students respond to this tension in authentic classrooms remains largely unexplored.

Navigating this paradox requires support grounded in evidence from students’ conceptual understanding and task-related activities, not dialogue alone. Multimodal learning analytics (MMLA) combines environment logs with other data sources to support inferences that chat histories cannot \citep{cohn2025theory,ouyang2026systematic,zhang2026using}, such as whether students can explain the concepts underlying their models and whether their reliance on the agent declines as their competence grows \citep{winne2017cognition}.

Agentic AI provides the technical means to act on this evidence \citep{stryker_agentic_ai_ibm}, but such flexibility requires pedagogical guardrails. We, therefore, adopt a stance of \textit{bounded autonomy} \citep{pan2025measuring}, where the agent formulates support within theory-derived constraints \citep{khosravi2022explainable}. These decisions occur within a broader classroom ecology. Students learn through interactions with partners, teachers, and tools \citep{song2021review}. Evaluating pedagogical agents requires considering how this support operates alongside external sources of assistance, not only within student-agent dialogue.

Prior LLM pedagogical-agent research has advanced on several fronts, but rarely in combination (see Section~\ref{sec:background}), and classroom research is still needed to examine how students' engagement with such agents relates to learning rather than task completion alone \citep{wu2025effects,ganguly2026conversational}. Because authentic deployments often preclude randomized designs, design-based studies provide an appropriate means of examining how these systems function in situ \citep{wang2005design,jung2023design}. We therefore frame the present study as an exploratory design case rather than an effectiveness trial.

In response to this need, we developed and deployed CoSMoAgent, an agentic, multi-agent, multimodal \textbf{Co}mputational \textbf{S}cience \textbf{Mo}deling \textbf{Agent} in C2STEM, an OELE in which students build computational models of kinematic phenomena \citep{hutchins2020c2stem}. CoSMoAgent implements the Evidence–Decision–Feedback (EDF) framework (\citealp{cohn2026edf}; see Section~\ref{sec:edf}), linking evidence of students’ model-building progress and strategies to pedagogical decisions and feedback. We examined its use in a high school science classroom with 11 student dyads across three modeling tasks, yielding 33 dyad-session observations.

We address three exploratory research questions. \textbf{RQ1} uses two contrasting focal dyads with similar normalized pre–post learning gains to examine how they differed in task mastery, agent-interaction frequency, peer collaboration, and problem-solving strategies, interpreted in light of their prior knowledge. \textbf{RQ2} examines how students’ verbalized understanding and reliance-oriented help seeking changed as task mastery increased. \textbf{RQ3} examines how CoSMoAgent’s dialogue-policy selection changed as task mastery evolved. Together, these questions distinguish task performance from learning gains and characterize learning processes and adaptive support during open-ended problem solving, without establishing causal effects.

This study uses the same classroom deployment previously reported in introducing the Evidence--Decision--Feedback (EDF) framework and its multi-agent architecture, here instantiated as CoSMoAgent \citep{cohn2026edf}: the same 11 dyads, three tasks, and C2STEM environment. RQ2 and RQ3 draw on this shared dataset and overlap with findings from that work, but are included here to contextualize RQ1 within the broader design case. RQ1 is new to the present study and relates task mastery, agent interaction, collaboration, and problem-solving strategy to independently measured pre--post learning gains, which were not examined previously.

The study makes four contributions. First, it presents CoSMoAgent's multi-agent architecture, including the temporal separation of asynchronous evidence reasoning from synchronous dialogue as an underexplored design dimension for LLM agents. Second, it reinforces the importance of distinguishing task completion from learning as a design and measurement consideration. Third, it provides qualitative evidence of how students respond to agent probing and how such interactions can redirect reasoning toward peer collaboration. Fourth, it identifies implementation considerations for answer-withholding agents, including students' expectations of direct answers and the corresponding importance of onboarding and teacher framing.

The remainder of the paper reviews related work (Section~\ref{sec:background}), presents the EDF framework (Section~\ref{sec:edf}) and CoSMoAgent architecture (Section~\ref{sec:copa}), describes the study methods (Section~\ref{sec:methods}), reports findings (Section~\ref{sec:findings}), and discusses implications, limitations, and future directions (Section~\ref{sec:discussion}).

\section{Background} \label{sec:background}

\subsection{Scaffolding in OELEs}

Before LLMs, intelligent tutoring system (ITS) frameworks such as \emph{Adaptive Control of Thought-Rational} (ACT-R) \citep{ritter2019act} and \emph{Knowledge-Learning-Instruction} (KLI) \citep{koedinger2012knowledge} modeled human cognition to align instruction with cognitive processes, enabling precise model-tracing feedback within a deliberately narrow problem-solving scope.

OELEs, in contrast, organize learning around less-constrained inquiry, exploration, experimentation, and problem solving through approaches such as ``learning by teaching'' \citep{munshi2023analysing} and ``learning by modeling'' \citep{hutchins2020c2stem,zhang2020studying}. This creates a different scaffolding challenge from traditional ITSs: support must respond to evolving problem-solving processes and multiple forms of learner evidence without overly constraining exploration. In STEM+C modeling, for example, students may produce functional code without understanding the underlying science or revise simulations without grasping the reasoning behind those changes. Thus, while ACT-R and KLI remain important foundations for theory-guided tutoring, OELEs require feedback that extends beyond correctness or task completion to support disciplinary reasoning, strategic thinking, and productive problem solving.

Scaffolding was originally defined as temporary support that enables learners to accomplish what they could not yet do alone \citep{wood1976role}. Effective scaffolding is contingent on the learner's state, fades as competence grows, and gradually transfers responsibility to the learner \citep{van2010scaffolding}. These properties are difficult to sustain in software, where supports are often fixed rather than diagnostic or adaptive \citep{puntambekar2005tools}. OELE research has addressed this through learner-activity-driven scaffolding: CTSiM and its successor, C2STEM, used adaptive supports to help students connect scientific concepts with agent-based computational models \citep{basu2015scaffolding,hutchins2020c2stem}, but relied on predefined triggers and templated feedback. LLMs offer a path toward more flexible, dialogic scaffolding while preserving contingency, fading, and transfer of responsibility.

\subsection{LLM Pedagogical Agents}

The proliferation of LLM-based pedagogical agents creates new opportunities for adaptive, dialogic support in OELEs while increasing the need for theory-grounded design. Recent systems use multi-agent architectures, multimodal learner modeling, and agentic retrieval-augmented generation (RAG) \citep{hou_llm-enhanced_2025,vatral2023theoretical,cohn2026edf,vatral2023prediction}, but technical sophistication alone does not ensure pedagogical value.

Multi-agent architectures typically distribute functions such as assessment, retrieval, planning, and dialogue across specialized agents \citep{hou_llm-enhanced_2025,wang_llm-powered_2025,zhang2025eduplanner}. Yet their temporal coordination remains underexamined: reasoning that delays the conversational turn can disrupt the responsive pacing required for dialogic scaffolding.

Designing effective pedagogical agents therefore requires attention not only to how support is generated, but also to how its educational value is evaluated. A long-standing finding in the learning sciences is that performance during instruction is an unreliable indicator of learning \citep{soderstrom2015learning}; students who struggle with complex problems before receiving instruction can even outperform those given direct guidance \citep{kapur2010productive,kim2026designing}. In OELEs, task completion and rubric progress cannot substitute for independently measured learning, particularly as LLM agents become increasingly capable of moving students efficiently toward correct task states.

Distinguishing performance from learning requires examining how students engage with support during problem solving. Help seeking is a self-regulatory skill that learners often use ineffectively \citep{aleven2003help}. ITS research has documented both help avoidance and help abuse, including repeatedly requesting hints until the answer is revealed \citep{baker2004off}. Conversational agents exacerbate this behavior, making students' reliance on them an important process indicator. Declining reliance as competence grows is consistent with scaffolding's transfer of responsibility \citep{van2010scaffolding}, whereas persistent or increasing reliance may signal cognitive offloading.

These concerns about reliance ultimately raise a broader question: whether and under what conditions LLM agents improve learning. Evidence is mixed. In a large field experiment, high-school students given unrestricted access to a GPT-4 assistant performed better during practice but worse on unassisted exams, whereas guardrails largely mitigated this harm \citep{bastani2025generative}. By contrast, a randomized undergraduate-physics study found that a carefully designed AI tutor produced larger gains than an active-learning lesson \citep{kestin2025ai}. Together, these findings suggest that learning effects depend strongly on pedagogical design and reinforce the need to distinguish learning from task performance and engagement.

Existing LLM pedagogical agents also reveal complementary design and evaluation gaps. Some emphasize theoretically motivated or technically sophisticated architectures but lack substantial classroom evaluation \citep{scholz2025partnering,li_edumas_2024,dai_agent4edu_2025}; others have been deployed in classrooms but rely primarily on chat histories rather than multimodal environmental evidence, limiting adaptive scaffolding and feedback personalization \citep{shi_educationq_2025,cohn2025theory}. Few studies combine theory-guided design, multimodal learner evidence, and authentic classroom deployment while examining students' interactions with agent support.

Table~\ref{tab:background} presents an illustrative (rather than systematic) synthesis of recent LLM pedagogical-agent studies, highlighting agent- and study-design features relevant to the current investigation. We identified candidate papers through a targeted review of 2024--2026 publications in major AI and AIED venues (e.g., AIED, LAK) and leading educational-technology journals, then selected papers that (a) deploy an LLM-based conversational or interactive agent intended to support student learning and (b) report at least one summarized dimension, prioritizing diversity of design over exhaustive coverage.

Each \cmark\ reflects what is explicitly stated in the cited paper, coded conservatively when reporting was ambiguous. We include \citealp{chu_llm-powered_2025} as a deliberate boundary case: its GradeOpt system uses multiple LLM agents---a grader, reflector, and refiner---to optimize grading guidelines. These agents interact only with each other, not with students, so it is coded as multi-agent while receiving \xmark\ on every student-facing dimension. Thus, a multi-agent architecture alone does not constitute an interactive pedagogical agent.

As Table~\ref{tab:background} shows, prior studies have advanced theoretical grounding, personalization, adaptability, multimodality, and classroom implementation, but few combine these features while examining learning processes in authentic open-ended environments. This study addresses that gap through CoSMoAgent, a theory-guided, multimodal, agentic system deployed in an authentic STEM+C OELE.

\begin{table}[htbp]
    \centering
    \begin{threeparttable}
        \small
        \setlength{\tabcolsep}{2pt}
        \renewcommand{\arraystretch}{1.15}
        \resizebox{0.94\textwidth}{!}{%
        \begin{tabular}{lcccccc|
            >{\centering\arraybackslash}p{1.45cm}
            >{\centering\arraybackslash}p{1.45cm}
            >{\centering\arraybackslash}p{1.45cm}}
            \toprule
            & \multicolumn{6}{c|}{\textbf{Agent-design features}}
            & \multicolumn{3}{c}{\makecell{\textbf{Study-design/}\\\textbf{analytic characteristics}}} \\
            \cmidrule(lr){2-7} \cmidrule(lr){8-10}
            Paper
            & \shortstack{Theo.\\Found.}
            & Agentic
            & \shortstack{Multi-\\Agent}
            & \shortstack{Multi-\\modality}
            & \shortstack{Personal-\\ization}
            & \shortstack{Adapt-\\ability}
            & \shortstack{Authentic\\Study}
            & \shortstack{Interaction\\Patterns}
            & \shortstack{Reliance} \\
            \midrule
            Aulia et al. \citeyearpar{aulia2025guiding}
            & \cmark & \xmark & \xmark & \cmark & \cmark & \cmark
            & \xmark & \xmark & \cmark \\
    
            Cohn et al. \citeyearpar{cohn2025theory}
            & \cmark & \xmark & \xmark & \xmark & \cmark & \cmark
            & \cmark & \xmark & \xmark \\
    
            Chu et al. \citeyearpar{chu_llm-powered_2025}
            & \xmark & \xmark & \cmark & \xmark & \xmark & \xmark
            & \xmark & \xmark & \xmark \\
    
            Dai et al. \citeyearpar{dai_agent4edu_2025}
            & \xmark & \cmark & \cmark & \xmark & \cmark & \xmark
            & \xmark & \xmark & \xmark \\
    
            Hou et al. \citeyearpar{hou_llm-enhanced_2025}
            & \cmark & \xmark & \cmark & \xmark & \xmark & \xmark
            & \cmark & \cmark & \xmark \\
    
            Jin et al. \citeyearpar{jin2025learning}
            & \cmark & \xmark & \xmark & \cmark & \xmark & \xmark
            & \cmark & \xmark & \xmark \\
    
            Li et al. \citeyearpar{li_edumas_2024}
            & \xmark & \xmark & \cmark & \xmark & \cmark & \cmark
            & \xmark & \xmark & \xmark \\
    
            Liu et al. \citeyearpar{liu2025llm}
            & \cmark & \xmark & \xmark & \xmark & \xmark & \cmark
            & \cmark & \xmark & \xmark \\
    
            Liu et al. \citeyearpar{liu2025one}
            & \cmark & \xmark & \cmark & \xmark & \cmark & \cmark
            & \xmark & \xmark & \xmark \\
    
            Scholz et al. \citeyearpar{scholz2025partnering}
            & \xmark & \cmark & \cmark & \xmark & \cmark & \cmark
            & \xmark & \xmark & \cmark \\
    
            Shi et al. \citeyearpar{shi_educationq_2025}
            & \cmark & \xmark & \cmark & \xmark & \xmark & \cmark
            & \xmark & \cmark & \xmark \\
    
            Sixu et al. \citeyearpar{sixu2024developing}
            & \xmark & \xmark & \xmark & \cmark & \xmark & \cmark
            & \cmark & \xmark & \xmark \\
    
            Sun et al. \citeyearpar{sun_multitutor_2025}
            & \xmark & \xmark & \cmark & \xmark & \xmark & \xmark
            & \xmark & \xmark & \xmark \\
    
            Wang et al. \citeyearpar{wang_llm-powered_2025}
            & \xmark & \xmark & \cmark & \cmark & \cmark & \cmark
            & \cmark & \cmark & \xmark \\
    
            Xi et al. \citeyearpar{xi2025investigating}
            & \cmark & \xmark & \xmark & \xmark & \cmark & \cmark
            & \cmark & \cmark & \xmark \\
    
            Zha et al. \citeyearpar{zha_mentigo_2025}
            & \xmark & \xmark & \xmark & \cmark & \cmark & \cmark
            & \xmark & \cmark & \xmark \\
    
            Zhang et al. \citeyearpar{zhang2025eduplanner}
            & \xmark & \xmark & \cmark & \xmark & \cmark & \xmark
            & \xmark & \xmark & \xmark \\
    
            \midrule
            Ours (CoSMoAgent)
            & \cmark & \cmark & \cmark & \cmark & \cmark & \cmark
            & \cmark & \cmark & \cmark \\
            \bottomrule
            \end{tabular}%
        }
        \caption{Illustrative synthesis of recent LLM pedagogical-agent research, organized by agent-design features and study-design/analytic characteristics.}
        \label{tab:background}
        \vspace{2pt}
        \begin{tablenotes}[flushleft]
            \tiny
            \item \textbf{Legend.} TRUE=\cmark, FALSE=\xmark.
            \item
            \setlength{\tabcolsep}{6pt}
            \renewcommand{\arraystretch}{1.15}
            \begin{tabular}{@{}p{0.31\textwidth}p{0.31\textwidth}p{0.31\textwidth}@{}}
                \underline{\textit{Theo. found.}}: explicit grounding in learning theory. &
                \underline{\textit{Agentic}}: explicitly agentic. &
                \underline{\textit{Multi-agent}}: explicit multi-agent architecture. \\

                \underline{\textit{Multimodality}}: includes non-textual inputs, such as logs or other environmental data. &
                \underline{\textit{Personalization}}: agent offers tailored interactions. &
                \underline{\textit{Adaptability}}: scaffolding evolves with learners. \\

                \underline{\textit{Authentic study}}: in-situ study with classroom students. &
                \underline{\textit{Interact. patt.}}: analyzes student-agent interaction patterns. &
                \underline{\textit{Reliance}}: considers student reliance on the agent. \\
            \end{tabular}
        \end{tablenotes}
    \end{threeparttable}
\end{table}

\FloatBarrier

\section{Theoretical Framework}
\label{sec:edf}

To address limitations of ITS-based frameworks, \citet{cohn2026edf} proposed EDF, a pedagogical-agent scaffolding framework designed to provide feedback that is theory-grounded, interpretable to stakeholders, personalized to students, and adaptive over time. The present study extends that work by examining a question that work left open: \textit{how agent support relates to what students learn, rather than how far they progress through a task}. We conceptualize learning in the OELE at two interdependent levels. The first is \textit{conceptual knowledge}: the physics and computing concepts students need to build computational models. The second is \textit{problem-solving skill}: the ability to translate that knowledge into computational constructs, test and revise models, and select productive strategies.

Neither level is guaranteed by task completion: students may produce a working model without understanding why it works and may also learn without fully completing the task \citep{land2000cognitive,soderstrom2015learning}. We therefore treat task mastery as evidence for calibrating scaffolding rather than as a measure of learning, examining it alongside independently measured learning gains and verbalized understanding. EDF is designed not to move students efficiently toward correct task states, but to provide contingent support that preserves the cognitive and collaborative processes through which learning occurs. The agent must, therefore, sustain engagement without providing so much help that students offload their reasoning.

To specify how support is delivered, EDF draws on Social Constructivism and Social Cognitive Theory (SCT). Social Constructivism motivates scaffolding within learners' ZPD through dialogue that elicits explanation, reflection, and negotiation of meaning rather than replacing students' reasoning \citep{vygotsky1978mind}. SCT complements this by emphasizing learner agency and self-regulation, with support intended to help students increasingly direct and evaluate their own problem solving \citep{bandura2001social}. EDF operationalizes these principles by using learner evidence to determine when and how support should change, then translating those decisions into feedback that advances rather than performs the learner's cognitive work \citep{stadler2024cognitive}.

\begin{center}
    \includegraphics[width=.85\linewidth]{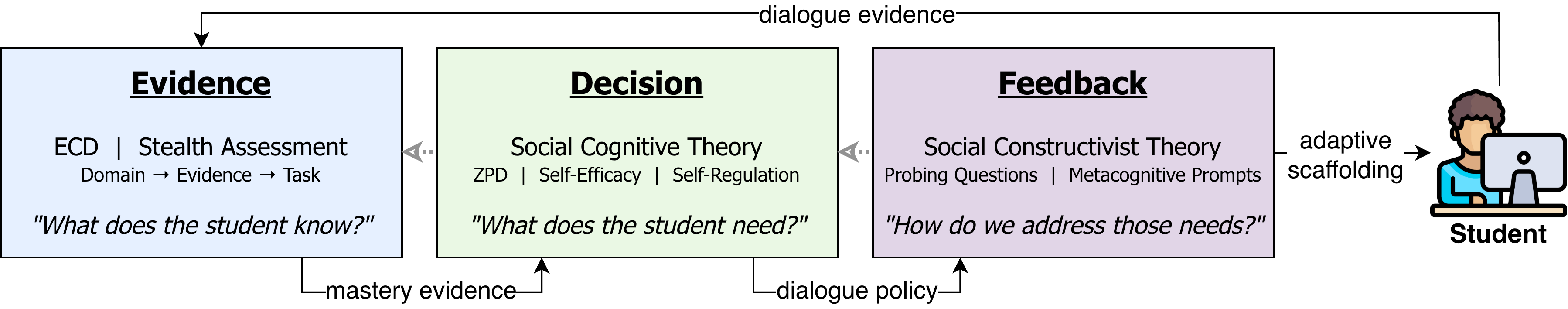}
    \captionof{figure}{Evidence-Decision-Feedback (EDF) framework \citep{cohn2026edf}.}
    \label{fig:edf}
\end{center}

As shown in Figure~\ref{fig:edf}, the \textbf{Evidence-Decision-Feedback (EDF)} framework comprises three semi-autonomous modules that jointly support adaptive scaffolding. The \textbf{Evidence} module, grounded in evidence-centered design (ECD) \citep{mislevy2003brief}, specifies the student data to monitor and updates the learner model accordingly. Drawing on Stealth Assessment \citep{shute2011stealth}, it unobtrusively assesses students' knowledge and metacognitive development as they progress through problem-solving tasks.

The \textbf{Decision} module uses \emph{mastery evidence} from the Evidence module to determine pedagogical intent and generate an SCT-aligned \emph{dialogue policy} within the student's ZPD. In doing so, it calibrates scaffolding to students' needs while supporting self-efficacy and self-regulation. The \textbf{Feedback} module then translates this policy into concrete conversational strategies grounded in social constructivist principles, such as active learning and reflection.

Gray dotted arrows in Figure~\ref{fig:edf} denote \emph{interpretability}, defined by \citet{cohn2026edf} as ``\textit{the extent to which agent output can be traced through the input-to-evidence-to-policy-to-feedback chain using observable system artifacts, thereby supporting stakeholders' ability to understand why the agent produced a given output}.'' CoSMoAgent preserves this property by logging the reasoning underlying each decision alongside the resulting policy and feedback. Interpretability is not evaluated as an outcome here; prior work found that these traces showed significantly above-baseline grounding in student data, alignment between evidence and decisions, and faithfulness between decisions and feedback \citep{cohn2026edf}.

\section{Computational Science Modeling Agent (CoSMoAgent)} \label{sec:copa}

\textbf{CoSMoAgent} is a GPT-5-powered, multi-agent, multimodal \textbf{Co}mputational \textbf{S}cience \textbf{Mo}deling \textbf{Agent} for STEM+C learning. It adopts a \emph{knowledgeable peer} persona, which students chose because it combines a peer's emotional support with an expert's competence \citep{cohn2025exploring}. This persona is intended to promote learner agency while reducing the dependence expert tutors can foster \citep{moos2009learning}. Rather than telling students what to do, CoSMoAgent asks questions, elicits explanations, and offers contingent hints that help students identify productive next steps while preserving ownership of the reasoning required to solve the problem.

CoSMoAgent operates agentically through coordinated sub-agents, but its autonomy is bounded by EDF. A rule-based tutor maps predefined states to predefined instructional actions, whereas an unconstrained general-purpose LLM may provide explanations or answers that bypass the learner's problem-solving process. CoSMoAgent occupies a middle ground: it reasons flexibly over learner evidence while EDF and its underlying learning theories constrain its goals, dialogue strategies, and feedback principles. This \emph{bounded autonomy} \citep{pan2025measuring} is maintained through human-in-the-loop design \citep{memarian2024human,alfredo2024human,fonteles2026jli,cohn2024towards} via iterative prompt refinement. \textit{Agent initiative}, in this sense, means deciding how to support a learner from available evidence, not what the learner should do.

CoSMoAgent is embedded in the \emph{C2STEM} learning environment (Figure~\ref{fig:c2stem}), a STEM+C OELE for learning one- and two-dimensional kinematics through computational modeling. Students work in dyads to simulate the motion of objects (e.g., a truck or drone) using kinematic equations and block-based code, then run their models and use graph and table tools to visualize changes in variables over time. They engage CoSMoAgent at their discretion on open-ended tasks with no single solution or problem-solving path.

\begin{center}
    \includegraphics[width=.85\linewidth]{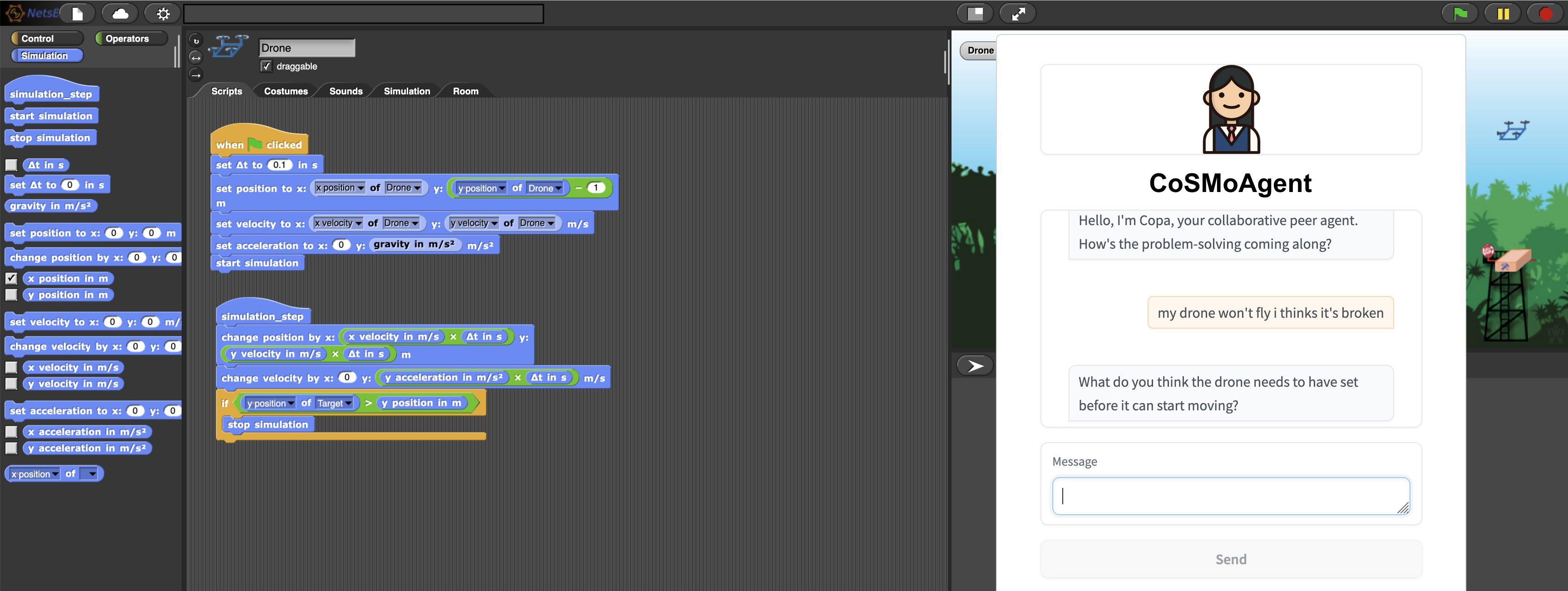}
    \captionof{figure}{CoSMoAgent, pictured within the C2STEM learning environment.}
    \label{fig:c2stem}
\end{center}

C2STEM interactions generate log data, including \emph{logged actions} (e.g., adding, editing, or removing blocks), \emph{model state} (the code blocks currently on screen), and \emph{task context} (e.g., variable initialization). Logged actions are transformed into semantically meaningful \emph{processed actions} for the LLM (e.g., \texttt{set\_block1234\_vel\_4} $\rightarrow$ \texttt{Set Velocity to 4 m/s}). Task progress is tracked using rubric criteria defined for each task, with students' models translated into a canonical form for evaluation.

\begin{center}
    \includegraphics[width=.85\linewidth]{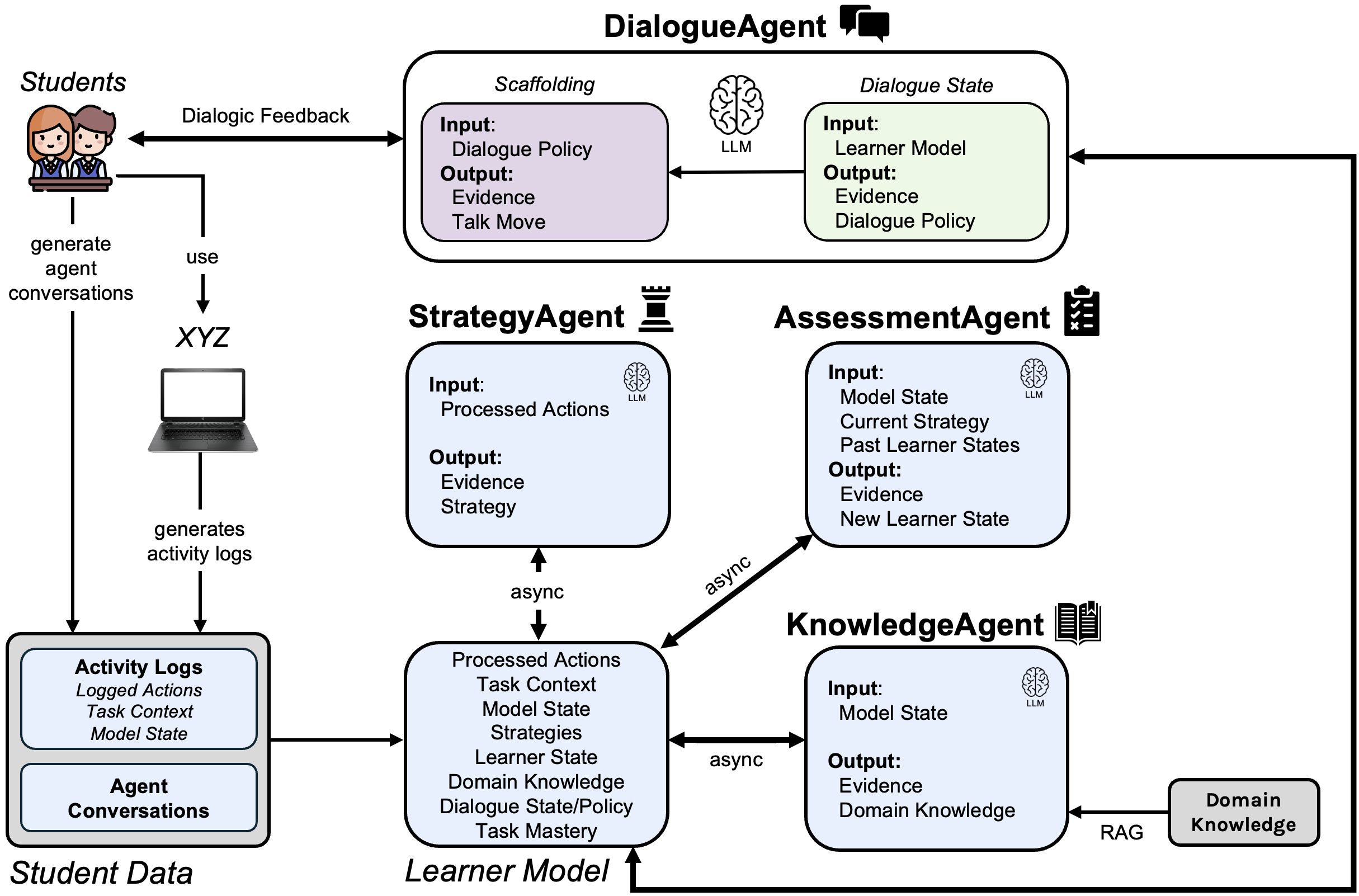}
    \captionof{figure}{CoSMoAgent and its four sub-agents. Colors correspond to the EDF modules presented in Figure~\ref{fig:edf}. Gray boxes = data stores. Thick edges = two-way relationships.}
    \label{fig:copa}
\end{center}

As shown in Figure~\ref{fig:copa}, CoSMoAgent comprises four specialized sub-agents operating over a shared \emph{learner model}. The learner model is continuously updated from \emph{activity logs} and \emph{agent conversations} and includes students' actions, model state, and task mastery. The sub-agents map onto the EDF framework: the \emph{Evidence} module (blue) contains three asynchronous agents---StrategyAgent, AssessmentAgent, and KnowledgeAgent---that transform activity data into a coherent representation of the learner.

The \textbf{StrategyAgent} reasons over students' last ten C2STEM actions to infer their current \emph{strategy} (e.g., \texttt{BUILD} $\rightarrow$ \texttt{ADJUST} $\rightarrow$ \texttt{EXECUTE} $\rightarrow$ \texttt{ADJUST}). It distinguishes productive strategies (e.g., \texttt{TINKERING}: systematic trial-and-error) from unproductive ones (e.g., \texttt{DEPTH\_FIRST\_ENACTING}: building without periodic testing), consistent with self-regulated learning research emphasizing temporal rather than isolated patterns of action \citep{snyder2024understanding}. These strategies capture \emph{how} students approach problem solving, not only \emph{what} they do on screen, and provide structured evidence for subsequent agent decisions.

The \textbf{AssessmentAgent} updates the \emph{learner state} (e.g., \texttt{DEBUGGING}, \texttt{PROGRESSING}) by jointly considering the current model state and recent strategy. For example, an error-free model built using a suboptimal strategy is classified as \texttt{EXPLORING}, whereas unresolved bugs after repeated actions indicate \texttt{STRUGGLING}. Consistent with Stealth Assessment, the agent unobtrusively evaluates performance to align subsequent decisions and feedback with students' current needs.

The \textbf{KnowledgeAgent} compares the current model state with an expert reference---a Python-like representation of a full-credit model---and uses the resulting gap to approximate students' ZPD. It then retrieves relevant \emph{domain knowledge} from a vector-store knowledge base derived from prior empirical accounts of student difficulties in C2STEM via an agentic RAG approach that uses environment log contextualization for retrieval (LC-RAG; \citealp{cohn2025personalizing}). Because retrieval is grounded in this model-state gap rather than semantic similarity to the student's query alone, the agent can identify relevant support even when students cannot clearly articulate their needs.

The \textbf{DialogueAgent} serves as the central orchestrator, integrating the \emph{Decision} (green) and \emph{Feedback} (purple) modules. These modules operate synchronously\footnote{Because student input is not known in advance, CoSMoAgent requires a query to infer a dialogue state, derive a dialogue policy, and formulate feedback.}, so combining them within one agent reduces response latency while the Evidence agents reason asynchronously. Upon receiving a query, the DialogueAgent infers the student's \emph{dialogue state} (e.g., \texttt{DEMONSTRATES\_UNDERSTANDING} or \texttt{INFORMATION\_SEEKING}) and integrates it with the learner model to determine pedagogical intent.

The resulting \emph{dialogue policy} specifies the agent's next pedagogical action: \texttt{PROBE\_UNDERSTANDING} when students seek direct solutions or express misconceptions, \texttt{SUGGEST\_ACTION} when a concrete environment action is warranted based on the students' knowledge, or \texttt{PUSH\_LIMIT} when students are ready to extend their reasoning beyond demonstrated task mastery. The policy is then realized through a \emph{talk move}, guided by instructions like ``\textit{promote self-efficacy through encouragement and praise}'' and ask ``\textit{probing questions}'' reflecting CoSMoAgent's social-cognitive and social-constructivist grounding.



All C2STEM task rubrics, agent prompts, LLM selection and prompt-engineering procedures, pre-/post-test materials, and detailed specifications of each agent's inputs, outputs, functionality, and theoretical grounding are provided in the \href{https://anonymous.4open.science/r/ETRD_Supplementary_Materials-6728/README.md}{Supplementary Materials}\footnote{\url{https://anonymous.4open.science/r/ETRD_Supplementary_Materials-6728/}}.

\section{Methods}
\label{sec:methods}

We examined students' learning outcomes and evaluated CoSMoAgent's role in supporting them by developing three research questions (RQs):

\begin{enumerate}[
    labelindent=\parindent,
    leftmargin=*]
    \item[RQ1.] \rqonea
    \item[RQ2.] \rqtwo
    \item[RQ3.] \rqthree
\end{enumerate}

\noindent Together, these questions examine the deployment at three levels: \textit{student learning outcomes}, \textit{student learning processes}, and \textit{agent adaptive behavior}. RQ1 examines whether task completion or engagement with CoSMoAgent reliably reflects learning, using two focal dyads with contrasting patterns of activity. RQ2 examines whether increasing task mastery is accompanied by stronger \textit{sense-making}, reflected in verbalized understanding, and reduced \textit{dependence} on the agent, reflected in reliance-oriented help seeking. RQ3 examines whether CoSMoAgent's \textit{dialogue-policy selection} shifts with task mastery as ZPD-based scaffolding would predict, from probing understanding to suggesting actions and extending reasoning.

Given the sample size and classroom context, the study is exploratory rather than experimental. We use a mixed-methods design: RQ1 is examined qualitatively through contrasting case studies, whereas RQ2 and RQ3 use descriptive and correlational analyses. These process measures characterize whether student--agent interactions align with CoSMoAgent's design goals; they are not used to infer causal effects on learning gains or independence.
 
\subsection{Participants and Setting}
\label{sec:participants}
 
In collaboration with five STEM educators, AIED researchers, and cognitive scientists (hereafter, ``the Educators''), we deployed CoSMoAgent in a high school science classroom in the southeastern United States. Participants were 11 sophomore dyads (ages 15--16; 48\% male, 48\% female, 2\% no response; 68\% White, 28\% Hispanic, 20\% Black, 12\% Asian, 4\% Native American)\footnote{Percentages exceed 100\% because some students identified as multiple ethnicities.} completing the six-week C2STEM kinematics curriculum. Weekly two-hour sessions were led by this paper's lead author with educator oversight and began with instruction connecting physics concepts (e.g., velocity and acceleration) to computational constructs (e.g., variables and loops), followed by modeling.

After two warm-up tasks to build familiarity with the agent, all dyads completed the three tasks analyzed here:
\begin{enumerate}[label=(\arabic*)]
  \item a 1-D Truck Task modeling a truck speeding up, cruising, slowing down, and stopping;
  \item a 2-D Drone Task programming a drone to drop a package on a target; and
  \item a 2-D Two-Package Drone Task programming a drone to drop two packages on separate targets.
\end{enumerate}
This yielded 33 dyad-session observations (11 dyads $\times$ 3 sessions).
 
\subsection{Data Sources}
\label{sec:data}
 
Across the 33 dyad-sessions, we collected the following:
\begin{itemize}
  \item 7,017 logged environment actions (mean $\approx$ 213 per dyad-session);
  \item 238 student--agent conversation turns (mean $\approx$ 7 per dyad-session); and
  \item the chain-of-thought reasoning associated with each agent decision (the ``evidence'' in Figure~\ref{fig:copa}).
\end{itemize}

Screen recordings, student video, and dyadic conversations were captured with webcams and lapel microphones using SyncFlow, an open-source multimodal data collection and alignment tool \citep{timalsina2025syncflow}. Students also completed pre- and post-tests and an anonymous exit survey. All findings were discussed with three participating Educators for additional feedback.

The same 32-point assessment, developed with the Educators who co-designed the C2STEM curriculum, was administered before and after the intervention. It included seven multi-part multiple-choice and short-answer items:
\begin{itemize}
  \item \textbf{Physics (four items, 16 points):} one- and two-dimensional kinematics concepts and their relationships.
  \item \textbf{Computing (three items, 16 points):} the computational constructs used to implement those concepts in C2STEM.
\end{itemize}

The assessment and survey are provided in the \href{https://anonymous.4open.science/r/ETRD_Supplementary_Materials-6728/README.md}{Supplementary Materials}.

\subsection{Measures}
\label{sec:measures}
 
\paragraph{Task mastery} is the percentage of rubric criteria satisfied by a dyad's model within a task, computed continuously from model state (Section~\ref{sec:copa}).

\paragraph{Learning gain} is the normalized gain $g = \tfrac{\text{post} - \text{pre}}{\text{max} - \text{pre}}$, computed from each student's pre- and post-test scores. Because students worked in dyads but were assessed individually, individual gains were averaged within dyads to align with dyad-level task mastery.

\paragraph{Engagement} is operationalized as agent-interaction frequency: the number of student--agent conversational turns initiated by a dyad within a session, summed across sessions for dyad-level comparisons.

\paragraph{Verbalized understanding} is the proportion of \texttt{PROBE\_UNDERSTANDING} interactions within a dyad-session followed by a \texttt{DEMONSTRATES\_UNDERSTANDING} dialogue state.

\paragraph{Reliance-oriented help seeking} is a dyad-session's student--agent interactions expressed as a proportion of the total student--agent interactions across all 33 dyad-sessions.
 
\subsection{Analysis}
\label{sec:analysis}
 
\paragraph{RQ1.} We selected two focal dyads with similar learning gains but contrasting patterns of task mastery and CoSMoAgent interaction. For each, we analyzed the full multimodal dataset (Section~\ref{sec:data}) across all three sessions. These cases are illustrative and not intended for population-level inference.
 
\paragraph{RQ2 and RQ3.} For each measure, we computed one value per dyad-session and calculated Spearman's $\rho$ across the resulting 33 dyad-session values, using each session's task-mastery bin as the ordinal predictor. RQ2 used deciles to capture fine-grained shifts in student behavior: reliance was calculated across all 10 deciles, whereas verbalized understanding used 7 (0--70\% mastery) because higher deciles contained too few \texttt{PROBE\_UNDERSTANDING} instances to estimate a success rate. RQ3 used quintiles corresponding to five meaningful phases of task progress, from initialization to the stopping condition. Policy frequencies were normalized within dyad-sessions to account for differences in interaction volume. Because the 33 dyad-sessions represent repeated observations of 11 dyads, these correlations are descriptive; no inferential analyses were conducted at the individual-student level.

\paragraph{Qualitative analysis.} All authors involved in data collection participated in the qualitative analysis, memoing key findings \citep{hatch2002,birks2008memoing} to detect and interpret patterns in the multimodal dataset. This included agent dialogue, peer discourse, screen activity, open-ended survey responses, and educator discussions. Interpretations were triangulated across these sources. These analyses supported RQ1, the illustrative episodes reported for RQ2 and RQ3, and the student and educator perspectives reported alongside them.

\paragraph{Survey.} Closed-ended Likert items are reported as descriptive means; we highlight the items most relevant to the study's central tension between theory-guided restraint and students' expectations of direct help. Open-ended responses were analyzed using the same memoing procedure described above.

\vspace{\baselineskip}
 
All participants and their parents provided informed assent and consent, and all procedures were approved by Vanderbilt University's Institutional Review Board.

\section{Findings} \label{sec:findings}


We treat task mastery, learning gains, and verbalized understanding as related but distinct measures (Section~\ref{sec:measures}). Accordingly, task progress does not necessarily indicate learning, and each RQ draws on these measures differently.

\subsection{RQ1: \textit{\rqonea}}
\label{sec:RQ1}

We examined RQ1 through the two focal dyads introduced in Section~\ref{sec:analysis}, across three sessions each (six dyad-session observations). The two dyads differed sharply in their use of CoSMoAgent despite completing approximately 900 C2STEM actions each. The low-engagement dyad, \emph{\LowEngage}, interacted with CoSMoAgent only 7 times across the three tasks (3, 1, and 3), primarily to confirm that their work was correct. Their prompts asked whether their code and model ``\textit{looked right},'' treating the agent largely as a validator rather than a problem-solving partner. CoSMoAgent responded with probes of understanding, which the students generally ignored. Yet low agent use did not reflect low task engagement: the dyad was highly collaborative, questioning one another, refining ideas together, and working beyond class time. Their infrequent use of CoSMoAgent may nevertheless have limited opportunities to benefit from its feedback during moments of uncertainty.

A representative example occurred during the Drone Task, when \emph{\LowEngage} asked, ``\textit{How do we know where to drop the package?}'' CoSMoAgent responded with a probe: ``\textit{What factors do you think decide how far the package travels before it hits the ground?}'' One student immediately rejected the response: ``\textit{That's not what I asked it at all!}'' This exchange reflects the dyad's persistent expectation of direct answers, even as they remained actively engaged with the task and with each other.

In contrast, the high-engagement dyad, \emph{\HighEngage}, produced 48 utterances (11, 22, and 15), eliciting a broader range of responses, including suggested actions and motivational prompts. They initially resembled \emph{\LowEngage} but increasingly treated CoSMoAgent as a problem-solving partner, using it for validation, causal questions, and expressions of frustration. Early exchanges during the Truck Task were primarily answer-seeking, and students did not consistently act on feedback. By the Drone Task, they more often read, discussed, and reasoned through the agent's hints, sometimes recognizing their own errors. By the Two-Package Drone Task, they primarily sought help getting ``\textit{in the right direction},'' and interactions more often followed a productive cycle of asking, discussing, enacting, and analyzing.

A clear example of this shift occurred on Day 1, when \emph{\HighEngage} reached an impasse and asked CoSMoAgent, ``\textit{we need help.}'' The agent acknowledged their difficulty and suggested involving the teacher: ``\textit{It looks like this part's tricky---maybe ask your teacher...}'' With teacher support, the students clarified their problem to CoSMoAgent (``\textit{it should stop going faster}''), prompting a more directive response: ``\textit{Nice! Try updating your condition so it checks when velocity is greater than the speed limit.}'' The dyad implemented the change, tested the model, and saw that it worked. This tangible progress appeared to shift how they engaged with CoSMoAgent, making them more willing to treat its feedback as a resource for subsequent reasoning rather than an obstacle to a direct answer.

\begin{figure}
    \centering
    \includegraphics[width=1\linewidth]{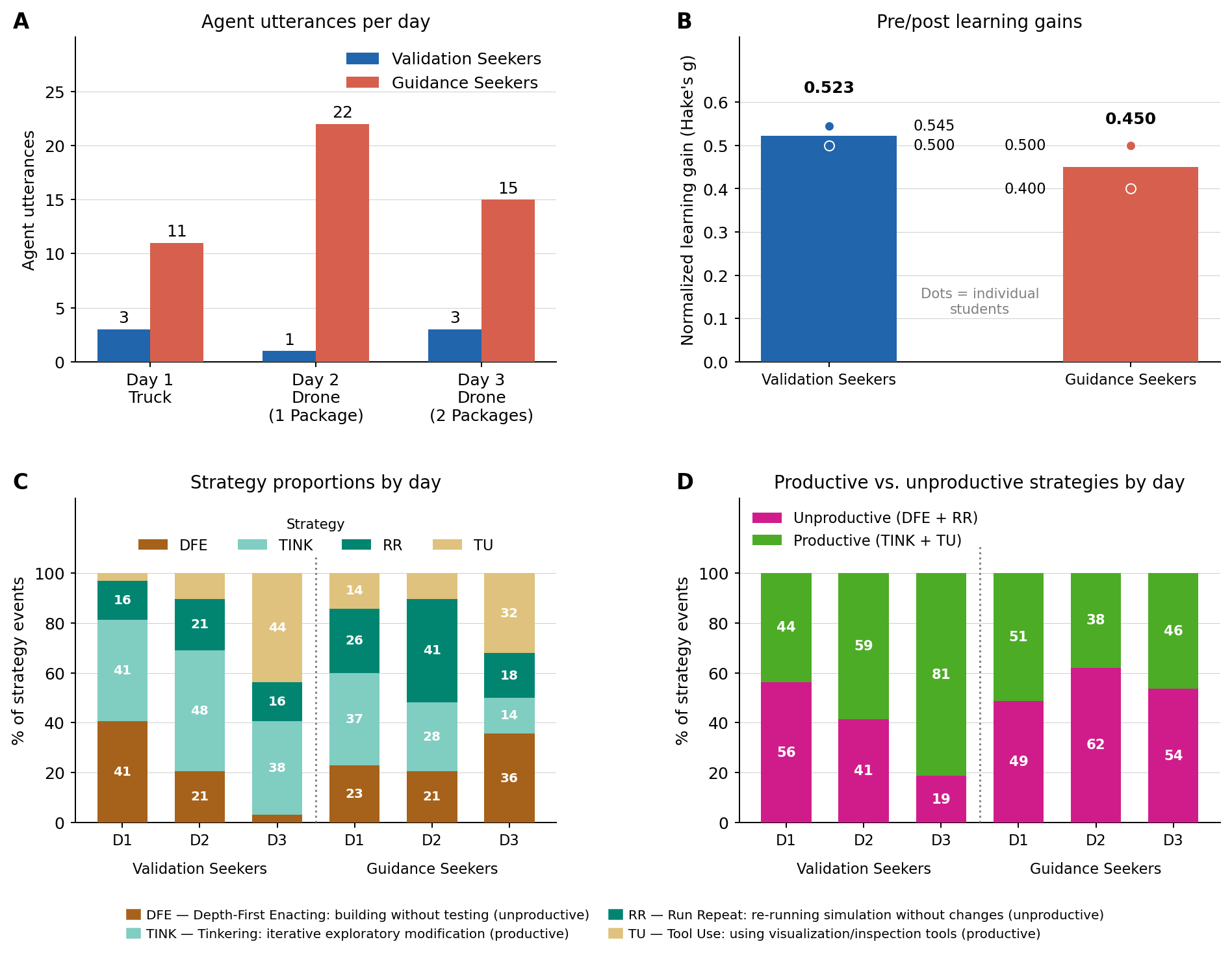}
    \caption{Contrasting trajectories for two focal dyads across three sessions. (A) Agent utterances. (B) Dyad-level normalized learning gain with individual gains shown as dots. (C) Problem-solving strategy proportions. (D) Productive versus unproductive strategies.}
    \label{fig:learning_cases}
\end{figure}

Figure~\ref{fig:learning_cases} compares the dyads' agent use, learning gains, and problem-solving strategies across the three sessions. Their strategy trajectories differed substantially (panels C--D): \emph{\LowEngage} shifted away from depth-first enacting toward more productive strategies, whereas \emph{\HighEngage} showed decreasing productive tinkering and increasing depth-first enacting. Video and audio data suggest the latter reflected a rapid, build-heavy approach in which students often followed agent guidance without testing between changes.

Despite these differences, \emph{\HighEngage} interacted with CoSMoAgent far more often than \emph{\LowEngage} (48 vs.\ 7 utterances) but showed a lower mean normalized learning gain (0.450 vs.\ 0.523; Figure~\ref{fig:learning_cases}A--B). The dyads also differed in prior knowledge: relative to the class-wide mean pretest score of 18.2/32, \emph{\LowEngage} entered above average (19.5/32) and \emph{\HighEngage} below it (16.0/32).

Together, these cases show that similar learning gains can emerge from markedly different problem-solving paths, including different levels of task mastery, agent use, prior knowledge, collaboration, and strategy. Thus, \emph{how} students engage with CoSMoAgent may be more informative than simply \emph{whether} or \emph{how often} they engage. RQ2 examines two additional aspects of this engagement: verbalized understanding and reliance-oriented help seeking.

\subsection{RQ2: \textit{\rqtwo}}

\subsubsection*{Verbalized understanding}

\begin{wrapfigure}[13]{r}{0.52\linewidth}
    \vspace{-5.0em}
    \centering
    \includegraphics[width=\linewidth]{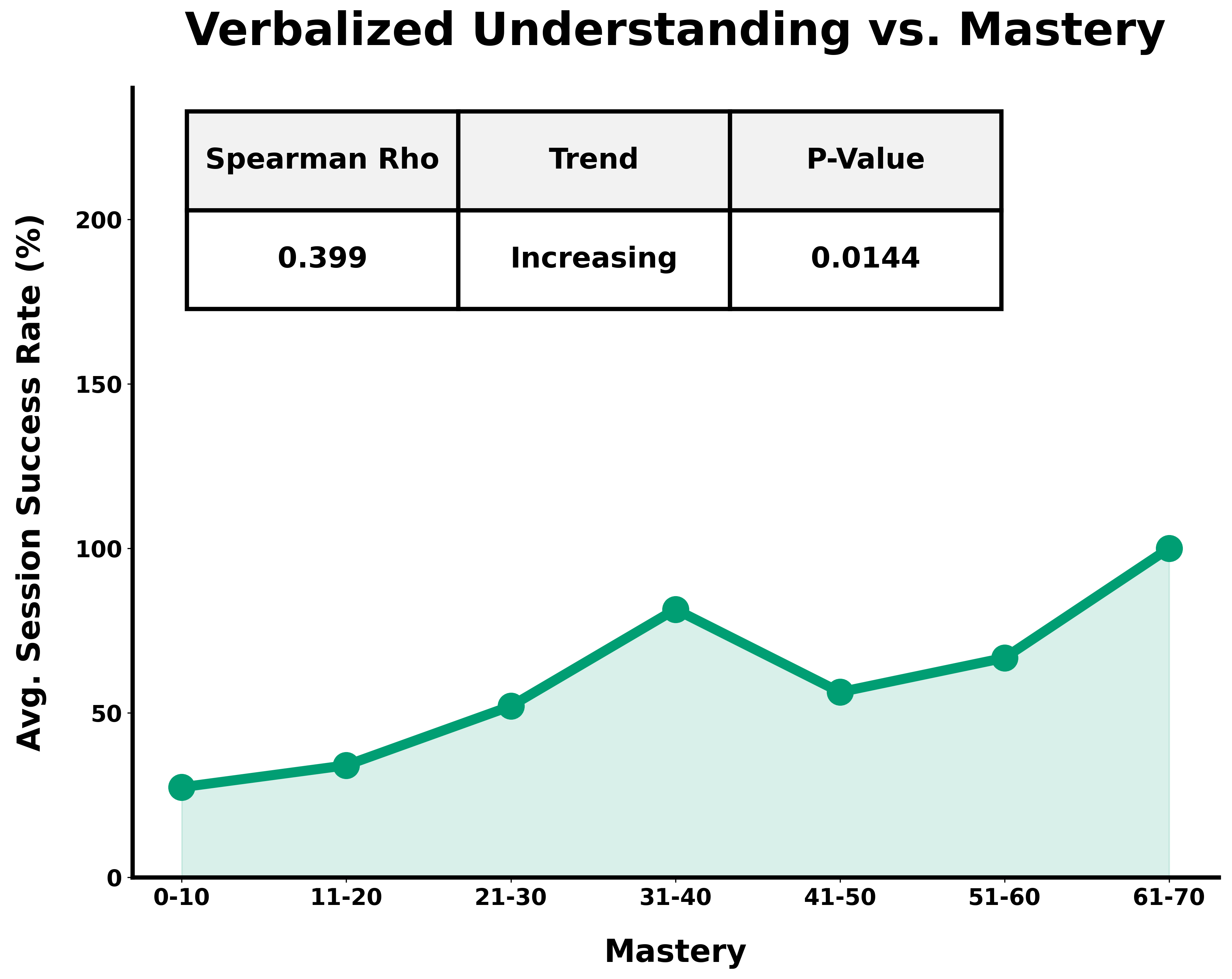}
    \caption{Verbalized-understanding success rate across task-mastery deciles.}
    \label{fig:understanding}
\end{wrapfigure}

Task performance alone cannot reveal whether students understand the concepts underlying their models. From a social constructivist perspective, such understanding is reflected in how students articulate and negotiate meaning with others. We therefore examined whether verbalized-understanding success following \texttt{PROBE\_UNDERSTANDING} policies varied across task-mastery deciles. As shown in Figure~\ref{fig:understanding}, success increased with mastery ($\rho = 0.40$, $p = 0.014$).\footnote{Success rates above $0.7$ mastery were not computed because \texttt{PROBE\_UNDERSTANDING} policies were infrequently applied for high mastery values; at high mastery, CoSMoAgent reduced probing as students no longer requested direct answers or exhibited misconceptions.} Thus, greater in-environment progress co-occurred with stronger verbal articulation of domain concepts during interaction with CoSMoAgent, consistent with conceptual sense-making rather than task completion alone.

For example, at the start of the Truck Task (0\% mastery), one dyad asked, ``\textit{Is our script correct so far?}'' hoping CoSMoAgent would identify errors in their model. Rather than answering directly, CoSMoAgent used a \texttt{PROBE\_UNDERSTANDING} policy: ``\textit{What other quantities do we need to set before the simulation can start correctly?}'' The students deflected the probe and again sought a direct answer: ``\textit{I meant our $\Delta t$, like is it supposed to be .1}''. Later in the same session, at 45\% mastery, CoSMoAgent asked, ``\textit{What quantities do you think connect how fast the truck is going to how far away it needs to start slowing down?}'' This time, the students replied ``\textit{acceleration and velocity},'' correctly identifying the relevant physics quantities and receiving a \texttt{DEMONSTRATES\_UNDERSTANDING} classification. This shift from deflection to articulation illustrates how greater task mastery coincided with stronger verbalization of the underlying physics.

\subsubsection*{Reliance-oriented help seeking}

\begin{wrapfigure}[18]{l}{0.52\linewidth}
    \centering
    \includegraphics[width=\linewidth]{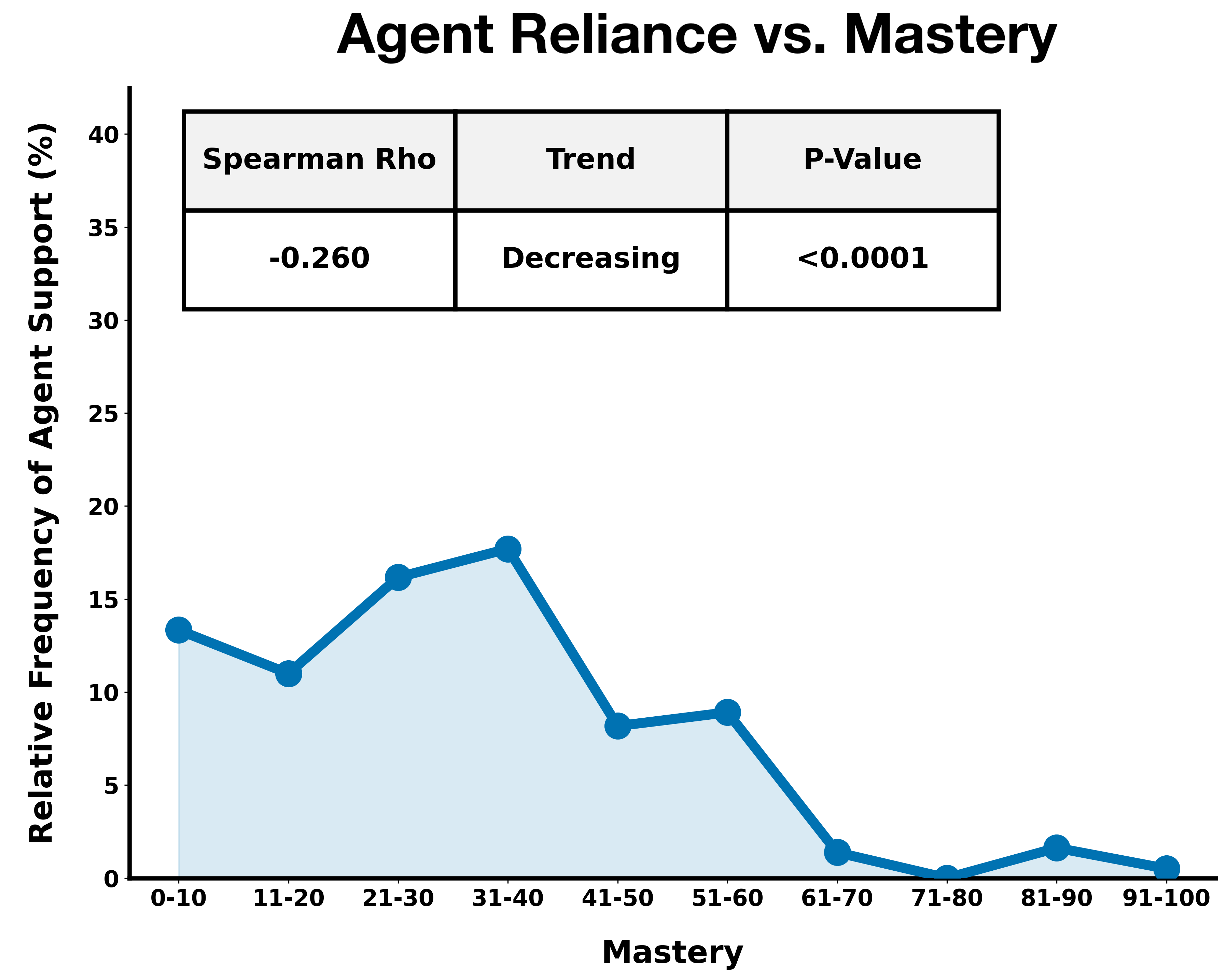}
    \caption{Reliance-oriented help seeking across task-mastery deciles.}
    \label{fig:reliance}
\end{wrapfigure}

During participatory design, students expressed a desire to control when and how to interact with CoSMoAgent \citep{cohn2025exploring}, raising the question of whether this agency would translate into sustained dependence. In this context, declining help-seeking behavior as mastery increases is consistent with SCT: students increasingly draw on their own knowledge and peer discussion rather than external support. Figure~\ref{fig:reliance} shows that 59\% of requests for agent support occurred below 40\% mastery. Task mastery was negatively associated with requests for agent support ($\rho = -0.26$, $p < 0.001$), indicating less help-seeking at higher mastery levels. Triangulation across student discourse, screen recordings, and video likewise showed greater use of peer discussion and independent reasoning as proficiency increased.

For example, one dyad illustrates this pattern during the Two-Package Drone Task. At 0\% mastery, a student asked, ``\textit{how should I calculate the y displacement equation?},'' seeking a direct answer. CoSMoAgent instead probed the dyad's understanding: ``\textit{What quantities do you think affect how far something moves vertically over time?}'' Rather than replying to the agent, the student turned to his partner: ``\textit{I think...gravity is how far something moves vertically over time}.'' Although incorrect, the response initiated peer discussion. The dyad then completed 288 consecutive actions---building, adjusting, testing, and using C2STEM's graph tool---without further agent support. They returned to CoSMoAgent only near completion for validation: ``\textit{We have made the packages drop at the right rate of descent and velocity. Is our code good now?}'' CoSMoAgent identified remaining problems in how package velocity and position were updated and again responded with a probe: ``\textit{How should the packages' velocity change once gravity starts acting on them?}''

Other dyads showed similar patterns, increasingly discussing CoSMoAgent's probes with partners or reasoning independently rather than seeking external help. These episodes suggest that probing, rather than direct answers, was associated with students reasoning on their own, consistent with SCT's emphasis on increasingly self-directed problem solving. We return to the broader implications of ``helpfulness'' in Section~\ref{sec:discussion}.

\subsection{RQ3: \textit{\rqthree}}

Consistent with ZPD-based scaffolding, CoSMoAgent should first probe students' understanding to identify gaps or misconceptions, then provide more specific guidance as mastery develops. We therefore examined whether dialogue-policy selection varied across task-mastery quintiles (Table~\ref{tab:adaptivity}).

\begin{table}[htbp]
    \centering
    \begin{tabular}{lccc}
        \hline
        Dialogue Policy & Spearman's $\rho$ & Trend & $p$-value \\
        \hline
        \texttt{PROBE\_UNDERSTANDING}   & $-0.34$ & Decreasing & $0.034$ \\
        \texttt{SUGGEST\_ACTION}        & $0.33$  & Increasing & $0.039$ \\
        \texttt{PUSH\_LIMIT}            & $0.42$  & Increasing & $0.007$ \\
        \hline
    \end{tabular}
    \caption{CoSMoAgent's dialogue policy adaptation across task mastery quintiles.}
    \label{tab:adaptivity}
\end{table}

As mastery increased, \texttt{PROBE\_UNDERSTANDING} decreased ($\rho = -0.34$, $p = 0.034$), whereas \texttt{SUGGEST\_ACTION} ($\rho = 0.33$, $p = 0.039$) and \texttt{PUSH\_LIMIT} ($\rho = 0.42$, $p = 0.007$) increased.

The following Truck Task interaction illustrates this progression. At 0\% mastery, a dyad reported, ``\textit{The truck isn't starting at -60 meters}.'' CoSMoAgent responded with \texttt{PROBE\_UNDERSTANDING}: ``\textit{What happens if the starting values are set after the simulation has already begun running?},'' targeting their understanding of initialization rather than correcting the error directly. At 65\% mastery, after the students reported that their model stopped correctly, CoSMoAgent shifted to \texttt{SUGGEST\_ACTION}: ``\textit{Want to try using the speed limit variable instead of the number for cruising?}'' After task completion, it used \texttt{PUSH\_LIMIT}: ``\textit{Your truck model works perfectly! Now let's tweak some values and see what happens!},'' encouraging further exploration.

This progression aligns with ZPD-based scaffolding: CoSMoAgent moved from probing what students understood, to suggesting concrete next steps, to extending their reasoning beyond demonstrated mastery. Overall, CoSMoAgent's scaffolding varied systematically with task mastery, shifting from diagnosis toward increasingly directive and extension-oriented support as students progressed.

\subsection{Student and Educator Perspectives on Theory-Guided Scaffolding}

Post-study survey responses reflected mixed views of CoSMoAgent. On a five-point Likert scale, students rated ``\textit{the questions CoSMoAgent asked were appropriate}'' at $M=3.81$ and ``\textit{asks questions that made me think about my model}'' at $M=3.88$. One student described a helpful interaction:

\begin{quote}
    When our model wasn't working great, and we had no clue why, CoSMoAgent asked specific questions that made us look deeply into the acceleration part of our code and fix the bug.
\end{quote}

At the same time, students rated ``\textit{I feel like CoSMoAgent understood what I was doing}'' at only $M=2.69$ and ``\textit{I feel like CoSMoAgent's suggestions were useful}'' at $M=2.77$. One student called CoSMoAgent ``\textit{broken},'' while another explained, ``\textit{One week we were very lost and asked CoSMoAgent to walk us through what to do, but all [CoSMoAgent] did was ask us questions.}'' These responses illustrate the tension between CoSMoAgent's scaffolded questioning and students' expectations for direct assistance.

Multimodal observations provided additional context. When students stopped seeking direct answers, they often shifted toward peer discussion, persistence, and independent reasoning, sometimes developing solutions through their own collaboration. Thus, the same answer-withholding behavior that frustrated students could also redirect problem solving away from the agent and toward their partners.

Educators described a similar tension in their classrooms. They noted students' frustration when answers were not immediately provided (``\textit{I agree with your conclusion...students wanting immediate answers}'') and linked these expectations to prior experiences with ChatGPT (``\textit{ChatGPT [is] so desperate to provide help...they've already become accustomed}''). One Educator summarized the resulting conflict between instructional goals and immediate student preferences: students often ``\textit{just [want]...the right answer...check the box and move on.}'' Educators also noted that students hold AI agents to different expectations than human instructors, underscoring the importance of onboarding and teacher framing to establish why CoSMoAgent withholds direct answers. We return to these implementation implications in Section~\ref{sec:discussion}.

\section{Discussion and Conclusions} \label{sec:discussion}

This study contributes to educational technology research and development by examining how learning theory can be operationalized in an LLM pedagogical agent across learner evidence, pedagogical decisions, and feedback while preserving students' ownership of open-ended problem solving.

The contrasting case-study dyads achieved similar learning gains despite markedly different task mastery, agent use, and problem-solving trajectories. This cautions against treating task completion or interaction frequency as proxies for learning: students may reach similar outcomes through different combinations of conceptual understanding, trial and error, peer collaboration, teacher support, and agent scaffolding. What appears to matter is not simply how much students engage with an agent, but how that engagement is integrated into their broader problem-solving process.

These findings also have implications for the design and architecture of adaptive feedback in OELEs. In computational modeling, students' progress cannot be understood from chat or task scores alone; effective support must account for model state, recent actions, and problem-solving strategy. CoSMoAgent operationalizes this through an EDF-based evidence-to-feedback pipeline in which multimodal learner evidence informs dialogue-policy selection. Supporting this pipeline also requires attention to \textit{when} different forms of reasoning occur. Whereas multi-agent systems commonly distribute functions across assessment, retrieval, planning, and dialogue \citep{hou_llm-enhanced_2025,li_edumas_2024,wang_llm-powered_2025,zhang2025eduplanner}, CoSMoAgent additionally separates asynchronous Evidence agents from a synchronous DialogueAgent. This \textit{temporal decomposition} enables more complex learner modeling without delaying the conversational exchange, preserving the responsive dialogue through which understanding is negotiated \citep{vygotsky1978mind}. As agent reasoning expands to richer retrieval, longer-horizon inference, and multimodal fusion, separating background reasoning from the dialogue loop may become increasingly important for avoiding the slow interaction cycles that constrained earlier ITSs \citep{ritter2019act,koedinger2012knowledge}.

Another implication is that an LLM agent's educational value depends on how students appropriate it within the broader classroom ecology, consistent with work on technology-supported classroom orchestration \citep{song2021review}. CoSMoAgent was only one source of support: students also relied on partners, teachers, task materials, and their own emerging strategies. In several cases, its probes redirected activity away from the agent and toward peer reasoning. The design question is therefore not simply whether students accept an agent's feedback, but whether that feedback helps organize productive activity across the learning environment.

This dynamic sharpens the \textit{helpfulness paradox} \citep{cohn2026edf}. CoSMoAgent's refusal to provide direct answers frustrated students, yet the same restraint could prompt further reasoning; reliance-oriented help seeking also declined as task mastery increased. LLM agents for OELEs should therefore not be optimized solely for student satisfaction or conversational smoothness, consistent with work on AI-supported productive failure \citep{kim2026designing}. Agents that are too indirect risk being ignored, whereas ones that are too helpful risk performing the reasoning students are meant to develop.

The focal dyads illustrate these two failure modes and reveal where CoSMoAgent's current adaptation falls short. For \emph{\HighEngage}, increasing reliance on \texttt{SUGGEST\_ACTION} coincided with declining productive tinkering and increasing depth-first enacting (Figure~\ref{fig:learning_cases}C--D), suggesting that concrete guidance may sometimes encourage building without sufficient testing (Section~\ref{sec:RQ1}). At the other extreme, repeated probing in response to \emph{\LowEngage}' validation requests failed to sustain engagement. Navigating the helpfulness paradox therefore requires adapting not only to task mastery, but also to prior knowledge, help-seeking intent, and signs of frustration, consistent with evidence that optimal guidance varies with learner expertise \citep{kalyuga2007expertise}. Effective scaffolding must therefore respond not only to what students know, but also to how they seek and use support.

This tension also makes implementation part of the design problem. Classroom use of pedagogical AI requires coordination among tool design, classroom routines, and teacher framing \citep{jung2023design}. Student agency is insufficient if learners interpret answer withholding as system failure rather than intentional scaffolding \citep{xing2026unveiling,yan2025beyond}; productive use depends on onboarding and teacher framing that make the agent's pedagogical role explicit by fostering student buy-in.

\vspace{1em}

\noindent These findings should be interpreted in light of three primary limitations that also motivate future design and empirical work:
\begin{itemize}
\item Institutional constraints precluded an RCT, so our claims are descriptive and correlational rather than causal. Nevertheless, non-randomized classroom and design-based studies can provide valuable evidence about how educational technologies function in authentic contexts \citep{wang2005design,kim2015development,deKoster2017concept}.
\item RQ1 and the strategy-trajectory comparisons rely on two focal-dyad case studies rather than inferential analyses across the full sample. These cases illustrate contrasting patterns within this classroom and should not be interpreted as evidence that task mastery, agent interaction, and learning gains are unrelated at the population level, or that the observed strategy trajectories generalize beyond these dyads.
\item RQ2 and RQ3 use 33 dyad-session observations, but these are repeated observations of the same 11 dyads rather than 33 independent cases. Their quantitative findings should therefore be interpreted within the scale and structure of this classroom sample.
\end{itemize}

Three directions for future work follow from these limitations. First, CoSMoAgent's design assumptions should be tested experimentally. Controlled comparisons of probing, graduated hints, and direct answers could test the helpfulness paradox using transfer and learning gains rather than satisfaction or task completion; comparisons with a chat-only agent could test the value of environmental evidence; and comparisons with a sequential architecture could test whether temporal decomposition preserves responsiveness. Second, larger studies should use multilevel models to account for repeated observations within dyads. Third, studies across classrooms, grade levels, domains, and LLMs are needed to distinguish the effects of the EDF design from those of a particular model and to examine whether answer-withholding support benefits learners with different levels of prior knowledge and backgrounds.

Taken together, this study positions CoSMoAgent as a design case for integrating theory-guided LLM agents into open-ended STEM+C learning environments. Its contributions include a multi-agent architecture, evidence that task progress and agent engagement should not be conflated with learning, illustrative accounts of student--agent interaction, and implementation insights for answer-withholding agents. More broadly, the study shows how learning theory can be operationalized across learner evidence, adaptive scaffolding, and student-facing dialogue while preserving the cognitive and collaborative work central to open-ended problem solving. For ETR\&D, the central design question is therefore not how to make an LLM maximally helpful, but how to make its assistance pedagogically productive: sufficiently responsive to support learners while constrained enough to preserve autonomy, collaboration, critical thinking, and ownership of sense-making.






%
%

\bibliographystyle{spbasic}      
\bibliography{references}   

@inproceedings{vatral2023prediction,
  title={Prediction of Students’ self-confidence using multimodal features in an experiential nurse training environment},
  author={Vatral, Caleb and Lee, Madison and Cohn, Clayton and Davalos, Eduardo and Levin, Daniel and Biswas, Gautam},
  booktitle={International Conference on Artificial Intelligence in Education},
  pages={266--271},
  year={2023},
  organization={Springer}
}

@inproceedings{snyder2024analyzing,
  title={Analyzing students collaborative problem-solving behaviors in synergistic STEM+ C learning},
  author={Snyder, Caitlin and Hutchins, Nicole M and Cohn, Clayton and Fonteles, Joyce Horn and Biswas, Gautam},
  booktitle={Proceedings of the 14th Learning Analytics and Knowledge Conference},
  pages={540--550},
  year={2024}
}

@inproceedings{cohn2024towards,
  title={Towards a human-in-the-loop LLM approach to collaborative discourse analysis},
  author={Cohn, Clayton and Snyder, Caitlin and Montenegro, Justin and Biswas, Gautam},
  booktitle={International Conference on Artificial Intelligence in Education},
  pages={11--19},
  year={2024},
  organization={Springer}
}

@inproceedings{cohn2025exploring,
  title={Exploring the Design of Pedagogical Agent Roles in Collaborative STEM+C Learning},
  author={Cohn, Clayton and Fonteles, Joyce Horn and Snyder, Caitlin and Srivastava, Namrata and T S, Ashwin and Campbell, Desmond and Montenegro, Justin and Biswas, Gautam},
  booktitle={Proceedings of the 18th International Conference on Computer-Supported Collaborative Learning-CSCL 2025, pp. 330-334},
  year={2025},
  organization={International Society of the Learning Sciences}
}

@inproceedings{cohn2025personalizing,
  title={Personalizing Student-Agent Interactions Using Log-Contextualized Retrieval Augmented Generation (RAG)},
  author={Cohn, Clayton and Rayala, Surya and Snyder, Caitlin and Fonteles, Joyce Horn and Jain, Shruti and Mohammed, Naveeduddin and Timalsina, Umesh and Burriss, Sarah and T S, Ashwin and Srivastava, Namrata and Deweese, Menton and Eeds, Angela and Biswas, Gautam},
  booktitle={International Conference on Artificial Intelligence in Education Workshop on Epistemics and Decision-Making in AI-Supported Education.},
  url={https://arxiv.org/abs/2505.17238},
  year={2025},
  organization={Springer}
}

@inproceedings{cohn2025theory,
  title={A Theory of Adaptive Scaffolding for LLM-Based Pedagogical Agents},
  author = {Cohn, Clayton and Rayala, Surya and Srivastava, Namrata and Fonteles, Joyce Horn and Jain, Shruti and Luo, Xinying and Mereddy, Divya and Mohammed, Naveeduddin and Biswas, Gautam},
  booktitle={Accepted to the AAAI Conference on Artificial Intelligence. In press},
  year={2025}
}

@InProceedings{cohn2026edf,
    author="Cohn, Clayton
    and Guo, Siyuan
    and Rayala, Surya
    and Wang, Hanchen David
    and Mohammed, Naveeduddin
    and Timalsina, Umesh
    and Jain, Shruti
    and Eeds, Angela
    and Deweese, Menton
    and Osborn Popp, Pamela J.
    and Stanton, Rebekah
    and Walker, Shakeera L.
    and Ma, Meiyi
    and Biswas, Gautam",
    editor="Blanchard, Emmanuel G.
    and Chen, Guanliang
    and Chi, Min
    and Isotani, Seiji",
    title="Evidence-Decision-Feedback: Theory-Driven Adaptive Scaffolding for LLM Agents",
    booktitle="Artificial Intelligence in Education",
    year="2027",
    publisher="Springer Nature Switzerland",
    address="Cham",
    pages="1--16",
    isbn="978-3-032-29744-0"
}

@article{fonteles2026jli,
    title = {Analyzing embodied learning in classroom settings: A human-in-the-loop AI approach for multimodal learning analytics},
    journal = {Learning and Instruction},
    volume = {103},
    pages = {102274},
    year = {2026},
    issn = {0959-4752},
    doi = {https://doi.org/10.1016/j.learninstruc.2025.102274},
    author = {Joyce Horn Fonteles and Clayton Cohn and Efrat Ayalon and Mengxi Zhou and Ashwin T.S. and Eduardo Davalos and Zhijian Li and Surya Rayala and Divya Mereddy and Austin Coursey and Shruti Jain and Yike Zhang and Noel Enyedy and Joshua Danish and Gautam Biswas},
}

@article{bandura2001social,
  title={Social cognitive theory: An agentic perspective},
  author={Bandura, Albert},
  journal={Annual review of psychology},
  volume={52},
  number={1},
  pages={1--26},
  year={2001},
  publisher={Annual Reviews 4139 El Camino Way, PO Box 10139, Palo Alto, CA 94303-0139, USA}
}

@article{kosmyna2025your,
  title={Your brain on ChatGPT: Accumulation of cognitive debt when using an AI assistant for essay writing task},
  author={Kosmyna, Nataliya and others},
  journal={arXiv:2506.08872},
  year={2025}
}

@inproceedings{stamper2024enhancing,
  title={Enhancing llm-based feedback: Insights from intelligent tutoring systems and the learning sciences},
  author={Stamper, John and Xiao, Ruiwei and Hou, Xinying},
  booktitle={International Conference on Artificial Intelligence in Education},
  pages={32--43},
  year={2024},
  organization={Springer}
}

@book{vygotsky1978mind,
  title={Mind in society: The development of higher psychological processes},
  author={Vygotsky, Lev S},
  volume={86},
  year={1978},
  publisher={Harvard university press}
}

@article{shute2011stealth,
  title={Stealth assessment in computer-based games to support learning},
  author={Shute, Valerie J},
  journal={Computer games and instruction},
  volume={55},
  number={2},
  pages={503--524},
  year={2011},
  publisher = {Information Age Publishing}
}

@article{mislevy2003brief,
  title={A brief introduction to evidence-centered design},
  author={Mislevy, Robert J and Almond, Russell G and Lukas, Janice F},
  journal={ETS Research Report Series},
  volume={2003},
  number={1},
  pages={i--29},
  year={2003},
  publisher={Wiley Online Library}
}

@article{ganguly2026conversational,
  title={Conversational AI agents in education: An umbrella review of current utilization, challenges, and future directions for ethical and responsible use},
  author={Ganguly, Amrita and Mehjabin, Nafisa and Malik, Aqdas and Johri, Aditya},
  journal={AI and Ethics},
  volume={6},
  number={1},
  pages={72},
  year={2026},
  publisher={Springer}
}

@misc{stryker_agentic_ai_ibm,
  author  = {Cole Stryker},
  title   = {What is agentic AI?},
  year    = {2025},
  url     = {https://www.ibm.com/think/topics/agentic-ai},
  urldate = {2025-10-31},
  publisher = {IBM Think},
  editor = {Staff Editor}
}

@inproceedings{chu_llm-powered_2025,
	title = {A {LLM}-{Powered} {Automatic} {Grading} {Framework} with {Human}-{Level} {Guidelines} {Optimization}},
	language = {en},
	urldate = {2025-10-14},
    booktitle = {Educational Data Mining},
	publisher = {International Educational Data Mining Society},
	author = {Chu, Yucheng and others},
	month = {Jun},
	year = {2025},
}

@inproceedings{hou_llm-enhanced_2025,
	address = {Cham},
	title = {An {LLM}-{Enhanced} {Multi}-agent {Architecture} for {Conversation}-{Based} {Assessment}},
	isbn = {978-3-031-98417-4},
	language = {en},
	booktitle = {Artificial {Intelligence} in {Education}},
	publisher = {Springer Nature Switzerland},
	author = {Hou, Xinying and others},
	year = {2025},
	pages = {119--134},
}

@inproceedings{zha_mentigo_2025,
	address = {New York, NY, USA},
	series = {{CHI} '25},
	title = {Mentigo: {An} {Intelligent} {Agent} for {Mentoring} {Students} in the {Creative} {Problem} {Solving} {Process}},
	isbn = {979-8-4007-1394-1},
	shorttitle = {Mentigo},
	urldate = {2025-10-16},
	booktitle = {Proceedings of the 2025 {CHI} {Conference} on {Human} {Factors} in {Computing} {Systems}},
	publisher = {Association for Computing Machinery},
	author = {Zha, Siyu and others},
	month = apr,
	year = {2025},
	pages = {1--22},
}

@inproceedings{dai_agent4edu_2025,
	address = {New York, NY, USA},
	series = {{ICAIE} '24},
	title = {{Agent4EDU}: {Advancing} {AI} for {Education} with {Agentic} {Workflows}},
	isbn = {979-8-4007-1269-2},
	shorttitle = {{Agent4EDU}},
	urldate = {2025-10-22},
	booktitle = {Proceedings of the 2024 3rd {International} {Conference} on {Artificial} {Intelligence} and {Education}},
	publisher = {Association for Computing Machinery},
	author = {Dai, Ling and others},
	month = apr,
	year = {2025},
}

@inproceedings{sun_multitutor_2025,
	title = {{MultiTutor}: {Collaborative} {LLM} {Agents} for {Multimodal} {Student} {Support}},
	shorttitle = {{MultiTutor}},
	language = {en},
	urldate = {2025-10-23},
	booktitle = {Proceedings of the {Innovation} and {Responsibility} in {AI}-{Supported} {Education} {Workshop}},
	publisher = {PMLR},
	author = {Sun, Edward and Tai, LeAnn},
	month = mar,
	year = {2025},
	note = {{ISSN}: 2640-3498},
	pages = {174--190},
}

@inproceedings{li_edumas_2024,
	title = {{EduMAS}: {A} {Novel} {LLM}-{Powered} {Multi}-{Agent} {Framework} for {Educational} {Support}},
	shorttitle = {{EduMAS}},
	urldate = {2025-10-24},
	booktitle = {2024 {IEEE} {International} {Conference} on {Big} {Data} ({BigData})},
	author = {Li, Qiaomu and Xie, Ying and Chakravarty, Sumit and Lee, Dabae},
	month = dec,
	year = {2024},
	note = {{ISSN}: 2573-2978},
	pages = {8309--8316},
}

@inproceedings{wang_llm-powered_2025,
	address = {New York, NY, USA},
	series = {{WWW} '25},
	title = {{LLM}-powered {Multi}-agent {Framework} for {Goal}-oriented {Learning} in {Intelligent} {Tutoring} {System}},
	isbn = {979-8-4007-1331-6},
	urldate = {2026-02-28},
	booktitle = {Companion {Proceedings} of the {ACM} on {Web} {Conference} 2025},
	publisher = {Association for Computing Machinery},
	author = {Wang, Tianfu and others},
	month = may,
	year = {2025},
}

@inproceedings{shi_educationq_2025,
	address = {Vienna, Austria},
	title = {{EducationQ}: {Evaluating} {LLMs}' {Teaching} {Capabilities} {Through} {Multi}-{Agent} {Dialogue} {Framework}},
	isbn = {979-8-89176-251-0},
	shorttitle = {{EducationQ}},
	urldate = {2026-02-28},
	booktitle = {Proceedings of the 63rd {Annual} {Meeting} of the {Association} for {Computational} {Linguistics} ({Volume} 1: {Long} {Papers})},
	publisher = {Association for Computational Linguistics},
	author = {Shi, Yao and others},
	month = jul,
	year = {2025},
}

@inproceedings{sixu2024developing,
  title={Developing an llm-empowered agent to enhance student collaborative learning through group discussion},
  author={Sixu, AN and Yu, YANG and Yunsi, MA and Guandong, XU and others},
  booktitle={International Conference on Computers in Education},
  year={2024},
  pages = {n/a}
}

@article{xi2025investigating,
  title={Investigating the effects of an LLM-based Socratic conversational agent on students’ academic performance and reflective thinking in higher education},
  author={Xi, Linjin and Zhang, Yi and Wang, Qiyun},
  journal={Computers \& Education},
  pages={105494},
  year={2025},
  publisher={Elsevier}
}

@inproceedings{jin2025learning,
  title={Learning by teaching: Enhancing music learning through llm-based teachable agents},
  author={Jin, Lingxi and Lin, Baicheng and Hong, Mengze and So, Hyo-Jeong and Zhang, Kun},
  booktitle={International Conference on Artificial Intelligence in Education},
  pages={148--155},
  year={2025},
  organization={Springer}
}

@article{liu2025llm,
  title={LLM-based pedagogical agent for ICU simulation instructor training: A quasi-experimental study},
  author={Liu, Jingbang and Chen, Ting and Li, Shan and Xia, Yeru and Zhu, Hong and Wu, Ruijuan and Cao, Qinli and Gong, Xiaoyan and Wu, Lili},
  journal={Nurse Education Today},
  pages={106901},
  year={2025},
  publisher={Elsevier}
}

@inproceedings{liu2025one,
  title={One size doesn't fit all: A personalized conversational tutoring agent for mathematics instruction},
  author={Liu, Ben and Zhang, Jihai and Lin, Fangquan and Jia, Xu and Peng, Min},
  booktitle={Companion Proceedings of the ACM on Web Conference 2025},
  pages={2401--2410},
  year={2025}
}

@inproceedings{scholz2025partnering,
  title={Partnering with AI: A pedagogical feedback system for LLM integration into programming education},
  author={Scholz, Niklas and others},
  booktitle={European Conference on Technology Enhanced Learning},
  year={2025},
  organization={Springer}
}

@article{zhang2025eduplanner,
  title={Eduplanner: Llm-based multi-agent systems for customized and intelligent instructional design},
  author={Zhang, Xueqiao and Zhang, Chao and Sun, Jianwen and Xiao, Jun and Yang, Yi and Luo, Yawei},
  journal={IEEE Transactions on Learning Technologies},
  year={2025},
  publisher={IEEE}
}

@inproceedings{aulia2025guiding,
  title={Guiding Self-Regulated Learning with an LLM-Based Pedagogical Chatbot},
  author={Aulia, Fariza Eka and others},
  booktitle={2025 International Seminar on Application for Technology of Information and Communication (iSemantic)},
  pages={505--511},
  year={2025},
  organization={IEEE}
}

@article{xing2026unveiling,
  title={Unveiling interaction patterns between students and generative AI teachable agent: Focusing on students' agency and AI agents' authority},
  author={Xing, Wanli and others},
  journal={British Journal of Educational Technology},
  year={2026},
  publisher={Wiley Online Library}
}

@article{yan2025beyond,
  title={Beyond efficiency: Empirical insights on generative AI's impact on cognition, metacognition and epistemic agency in learning},
  author={Yan, Lixiang and Pammer-Schindler, Viktoria and Mills, Caitlin and Nguyen, Andy and Ga{\v{s}}evi{\'c}, Dragan},
  journal={British Journal of Educational Technology},
  volume={56},
  number={5},
  pages={1675--1685},
  year={2025},
  publisher={Wiley Online Library}
}

@article{hutchins2020c2stem,
  title={C2STEM: A system for synergistic learning of physics and computational thinking},
  author={Hutchins, Nicole M and Biswas, Gautam and Mar{\'o}ti, Mikl{\'o}s and L{\'e}deczi, {\'A}kos and Grover, Shuchi and Wolf, Rachel and Blair, Kristen Pilner and Chin, Doris and Conlin, Luke and Basu, Satabdi and others},
  journal={Journal of Science Education and Technology},
  volume={29},
  number={1},
  pages={83--100},
  year={2020},
  publisher={Springer}
}

@article{pan2025measuring,
  title={Measuring Agents in Production},
  author={Pan, Melissa Z and Arabzadeh, Negar and others},
  journal={arXiv preprint arXiv:2512.04123},
  year={2025}
}

@article{sengupta2013integrating,
  title={Integrating computational thinking with K-12 science education using agent-based computation: A theoretical framework},
  author={Sengupta, Pratim and Kinnebrew, John S and Basu, Satabdi and Biswas, Gautam and Clark, Douglas},
  journal={Education and information technologies},
  volume={18},
  number={2},
  pages={351--380},
  year={2013},
  publisher={Springer}
}

@article{weintrop2016defining,
  title={Defining computational thinking for mathematics and science classrooms},
  author={Weintrop, David and Beheshti, Elham and Horn, Michael and Orton, Kai and Jona, Kemi and Trouille, Laura and Wilensky, Uri},
  journal={Journal of science education and technology},
  volume={25},
  number={1},
  pages={127--147},
  year={2016},
  publisher={Springer}
}

@article{basu2015scaffolding,
  title={A scaffolding framework to support learning of emergent phenomena using multi-agent-based simulation environments},
  author={Basu, Satabdi and Sengupta, Pratim and Biswas, Gautam},
  journal={Research in Science Education},
  volume={45},
  number={2},
  pages={293--324},
  year={2015},
  publisher={Springer}
}

@article{stadler2024cognitive,
  title={Cognitive ease at a cost: LLMs reduce mental effort but compromise depth in student scientific inquiry},
  author={Stadler, Matthias and Bannert, Maria and Sailer, Michael},
  journal={Computers in Human Behavior},
  volume={160},
  pages={108386},
  year={2024},
  publisher={Elsevier}
}

@article{wood1976role,
  title={The role of tutoring in problem solving},
  author={Wood, David and Bruner, Jerome S and Ross, Gail},
  journal={Journal of child psychology and psychiatry},
  volume={17},
  number={2},
  pages={89--100},
  year={1976},
  publisher={Blackwell Publishing Ltd Oxford, UK}
}

@article{van2010scaffolding,
  title={Scaffolding in teacher--student interaction: A decade of research},
  author={Van de Pol, Janneke and Volman, Monique and Beishuizen, Jos},
  journal={Educational psychology review},
  volume={22},
  number={3},
  pages={271--296},
  year={2010},
  publisher={Springer}
}

@article{puntambekar2005tools,
  title={Tools for scaffolding students in a complex learning environment: What have we gained and what have we missed?},
  author={Puntambekar, Sadhana and Hubscher, Roland},
  journal={Educational psychologist},
  volume={40},
  number={1},
  pages={1--12},
  year={2005},
  publisher={Taylor \& Francis}
}

@article{soderstrom2015learning,
  title={Learning versus performance: An integrative review},
  author={Soderstrom, Nicholas C and Bjork, Robert A},
  journal={Perspectives on Psychological Science},
  volume={10},
  number={2},
  pages={176--199},
  year={2015},
  publisher={Sage Publications Sage CA: Los Angeles, CA}
}

@article{kapur2010productive,
  title={Productive failure in mathematical problem solving},
  author={Kapur, Manu},
  journal={Instructional science},
  volume={38},
  number={6},
  pages={523--550},
  year={2010},
  publisher={Springer}
}

@article{aleven2003help,
  title={Help seeking and help design in interactive learning environments},
  author={Aleven, Vincent and Stahl, Elmar and Schworm, Silke and Fischer, Frank and Wallace, Raven},
  journal={Review of educational research},
  volume={73},
  number={3},
  pages={277--320},
  year={2003},
  publisher={Sage Publications Sage CA: Thousand Oaks, CA}
}

@article{bastani2025generative,
  title={Generative AI without guardrails can harm learning: Evidence from high school mathematics},
  author={Bastani, Hamsa and Bastani, Osbert and Sungu, Alp and Ge, Haosen and Kabakc{\i}, {\"O}zge and Mariman, Rei},
  journal={Proceedings of the National Academy of Sciences},
  volume={122},
  number={26},
  pages={e2422633122},
  year={2025},
  publisher={National Academy of Sciences}
}

@article{kestin2025ai,
  title={AI tutoring outperforms in-class active learning: an RCT introducing a novel research-based design in an authentic educational setting},
  author={Kestin, Greg and Miller, Kelly and Klales, Anna and Milbourne, Timothy and Ponti, Gregorio},
  journal={Scientific Reports},
  volume={15},
  number={1},
  pages={17458},
  year={2025},
  publisher={Nature Publishing Group UK London}
}

@article{kalyuga2007expertise,
  title={Expertise reversal effect and its implications for learner-tailored instruction},
  author={Kalyuga, Slava},
  journal={Educational psychology review},
  volume={19},
  number={4},
  pages={509--539},
  year={2007},
  publisher={Springer}
}

@article{ritter2019act,
  title={ACT-R: A cognitive architecture for modeling cognition},
  author={Ritter, Frank E and Tehranchi, Farnaz and Oury, Jacob D},
  journal={Wiley Interdisciplinary Reviews: Cognitive Science},
  volume={10},
  number={3},
  pages={e1488},
  year={2019},
  publisher={Wiley Online Library}
}

@article{koedinger2012knowledge,
  title={The Knowledge-Learning-Instruction framework: Bridging the science-practice chasm to enhance robust student learning},
  author={Koedinger, Kenneth R and Corbett, Albert T and Perfetti, Charles},
  journal={Cognitive science},
  volume={36},
  number={5},
  pages={757--798},
  year={2012},
  publisher={Wiley Online Library}
}

@article{munshi2023analysing,
  title={Analysing adaptive scaffolds that help students develop self-regulated learning behaviours},
  author={Munshi, Anabil and Biswas, Gautam and Baker, Ryan and Ocumpaugh, Jaclyn and Hutt, Stephen and Paquette, Luc},
  journal={Journal of Computer Assisted Learning},
  volume={39},
  number={2},
  pages={351--368},
  year={2023},
  publisher={Wiley Online Library}
}

@phdthesis{snyder2024understanding,
  title={Understanding Students' Collaborative Problem Solving during STEM+ C Learning using Multimodal Analysis},
  author={Snyder, Caitlin},
  year={2024},
  school={Vanderbilt University}
}

@inproceedings{timalsina2025syncflow,
  title={SyncFlow: A Scalable Platform for Multimodal Learning Analytics},
  author={Timalsina, Umesh and Davalos\_Anaya, Eduardo and Sanda, Nihar and Zhang, Yike and Horn\_Fonteles, Joyce and T\_S, Ashwin and Biswas, Gautam},
  year={2025},
  publisher={Zenodo},
  booktitle={US Research Software Engineering Conference (USRSE25)},
  pages={n/a}
}

@book{hatch2002, 
    title = {Doing qualitative research in education settings}, 
    publisher = {SUNY Press}, 
    year = {2002}, 
    author = {Hatch, J. Amos}, 
    location={Albany, NY} 
}

@inproceedings{baker2004off,
  title={Off-task behavior in the cognitive tutor classroom: When students" game the system"},
  author={Baker, Ryan Shaun and others},
  booktitle={Proceedings of the SIGCHI conference on Human factors in computing systems},
  year={2004}
}

@article{moos2009learning,
  title={Learning with computer-based learning environments: A literature review of computer self-efficacy},
  author={Moos, Daniel C and Azevedo, Roger},
  journal={Review of educational research},
  volume={79},
  number={2},
  pages={576--600},
  year={2009},
  publisher={Sage Publications Sage CA: Los Angeles, CA}
}

@incollection{winne2017cognition,
  title={Cognition and metacognition within self-regulated learning},
  author={Winne, Philip H},
  booktitle={Handbook of self-regulation of learning and performance},
  pages={36--48},
  year={2017},
  publisher={Routledge}
}

@article{birks2008memoing,
  title={Memoing in qualitative research: Probing data and processes},
  author={Birks, Melanie and Chapman, Ysanne and Francis, Karen},
  journal={Journal of research in nursing},
  volume={13},
  number={1},
  pages={68--75},
  year={2008},
  publisher={Sage Publications Sage UK: London, England}
}

@article{wu2025effects,
  title={The Effects of GAI-Enhanced Pedagogical Agents in the Metaverse (GPAiM) on Elementary School Students’ Conceptual Understanding and Cognitive Engagement Patterns},
  author={Wu, Tinghui and Zhai, Xuesong and Song, Yanjie},
  journal={Computers \& Education},
  year={2025},
  publisher={Elsevier}
}

@article{ouyang2026systematic,
  title={A systematic review of multimodal learning analytics in computer-supported collaborative learning},
  author={Ouyang, Fan and Bai, Xianping},
  journal={Computers \& Education},
  year={2026},
  publisher={Elsevier}
}

@inproceedings{zhang2026using,
  title={Using Large Language Models to Detect Socially Shared Regulation of Collaborative Learning},
  author={Zhang, Jiayi and others},
  booktitle={Proceedings of the LAK26: 16th International Learning Analytics and Knowledge Conference},
  pages={883--890},
  year={2026}
}

@article{khosravi2022explainable,
  title={Explainable artificial intelligence in education},
  author={Khosravi, Hassan and Shum, Simon Buckingham and Chen, Guanliang and Conati, Cristina and Tsai, Yi-Shan and Kay, Judy and Knight, Simon and Martinez-Maldonado, Roberto and Sadiq, Shazia and Ga{\v{s}}evi{\'c}, Dragan},
  journal={Computers and education: artificial intelligence},
  volume={3},
  pages={100074},
  year={2022},
  publisher={Elsevier}
}

@inproceedings{zhang2020studying,
  title={Studying the interactions between science, engineering, and computational thinking in a learning-by-modeling environment},
  author={Zhang, Ningyu and Biswas, Gautam and McElhaney, Kevin W and Basu, Satabdi and McBride, Elizabeth and Chiu, Jennifer L},
  booktitle={International conference on artificial intelligence in education},
  pages={598--609},
  year={2020},
  organization={Springer}
}

@inproceedings{vatral2023theoretical,
  title={A Theoretical Framework for Multimodal Learner Modeling and Performance Analysis in Experiential Learning Environments.},
  author={Vatral, Caleb and Biswas, Gautam and Goldberg, Benjamin},
  booktitle={AI-GEL@ AIED},
  pages={68--78},
  year={2023}
}

@article{memarian2024human,
  title={Human-in-the-loop in artificial intelligence in education: A review and entity-relationship (ER) analysis},
  author={Memarian, Bahar and Doleck, Tenzin},
  journal={Computers in Human Behavior: Artificial Humans},
  volume={2},
  number={1},
  pages={100053},
  year={2024},
  publisher={Elsevier}
}

@article{alfredo2024human,
  title={Human-centred learning analytics and AI in education: A systematic literature review},
  author={Alfredo, Riordan and Echeverria, Vanessa and Jin, Yueqiao and Yan, Lixiang and Swiecki, Zachari and Ga{\v{s}}evi{\'c}, Dragan and Martinez-Maldonado, Roberto},
  journal={Computers and Education: Artificial Intelligence},
  volume={6},
  pages={100215},
  year={2024},
  publisher={Elsevier}
}

@article{kim2015development,
  title={Development of a design-based learning curriculum through design-based research for a technology-enabled science classroom},
  author={Kim, Paul and Suh, Esther and Song, Donggil},
  journal={Educational Technology Research and Development},
  volume={63},
  number={4},
  pages={575--602},
  year={2015},
  publisher={Springer}
}

@article{wang2005design,
  title={Design-based research and technology-enhanced learning environments},
  author={Wang, Feng and Hannafin, Michael J},
  journal={Educational technology research and development},
  volume={53},
  number={4},
  pages={5--23},
  year={2005},
  publisher={Springer}
}

@article{deKoster2017concept,
  title={Concept-guided development of technology in ‘traditional’and ‘innovative’schools: Quantitative and qualitative differences in technology integration},
  author={De Koster, Sandra and Volman, Monique and Kuiper, Els},
  journal={Educational Technology Research and Development},
  volume={65},
  number={5},
  pages={1325--1344},
  year={2017},
  publisher={Springer}
}

@article{song2021review,
  title={A review of how class orchestration with technology has been conducted for pedagogical practices},
  author={Song, Yanjie},
  journal={Educational Technology Research and Development},
  volume={69},
  number={3},
  pages={1477--1503},
  year={2021},
  publisher={Springer}
}

@article{kim2026designing,
  title={Designing AI systems to support a productive-failure-based learning: insights from adult learners on AI applications and AI system design principles},
  author={Kim, Jinhee and Lin, Xi and Yu, Seongryeong and Detrick, Rita},
  journal={Educational technology research and development},
  pages={1--35},
  year={2026},
  publisher={Springer}
}

@article{jung2023design,
  title={Design-based implementation research: milestones and trade-offs in designing a collaborative representation tool for engineering classrooms.},
  author={Jung, Jiyoon and Mercier, Emma},
  journal={Educational Technology Research \& Development},
  volume={71},
  number={6},
  pages={2457},
  year={2023}
}

@article{land2000cognitive,
  title={Cognitive requirements for learning with open-ended learning environments},
  author={Land, Susan M},
  journal={Educational Technology Research and Development},
  volume={48},
  number={3},
  pages={61--78},
  year={2000},
  publisher={Springer}
}


\end{document}